\documentclass[twocolumn]{aastex61}

\received{date}
\revised{date}
\accepted{date}
\submitjournal{ApJ}

\shorttitle{Extinction related brightness variations of V582 Aurigae}
\shortauthors{\'Abrah\'am, P., K\'osp\'al, \'A., Kun, M. et al.}

\begin{document}
	
\title{An UXor among FUors: extinction related brightness variations of the young eruptive star V582 Aur}

\author{P. \'Abrah\'am} \affiliation{Konkoly Observatory, Research
  Centre for Astronomy and Earth Sciences, Hungarian Academy of
  Sciences, Konkoly-Thege Mikl\'os \'ut 15-17, 1121 Budapest, Hungary} 
  \email{abraham@konkoly.hu}

\author{\'A. K\'osp\'al} \affiliation{Konkoly Observatory, Research
  Centre for Astronomy and Earth Sciences, Hungarian Academy of
  Sciences, Konkoly-Thege Mikl\'os \'ut 15-17, 1121 Budapest, Hungary}
  \affiliation{Max Planck Institute for Astronomy, K\"onigstuhl 17,
  69117 Heidelberg, Germany} 

\author{M. Kun} \affiliation{Konkoly Observatory, Research
  Centre for Astronomy and Earth Sciences, Hungarian Academy of
  Sciences, Konkoly-Thege Mikl\'os \'ut 15-17, 1121 Budapest, Hungary}

\author{O. Feh\'er} \affiliation{Konkoly Observatory, Research Centre
  for Astronomy and Earth Sciences, Hungarian Academy of Sciences,
  Konkoly-Thege Mikl\'os \'ut 15-17, 1121 Budapest, Hungary}
\affiliation{E\"otv\"os Lor\'and University, Department of Astronomy, P\'azm\'any P\'eter s\'et\'any 1/A, 1117 Budapest, Hungary}

\author{G. Zsidi} \affiliation{Konkoly Observatory, Research Centre
  for Astronomy and Earth Sciences, Hungarian Academy of Sciences,
  Konkoly-Thege Mikl\'os \'ut 15-17, 1121 Budapest, Hungary} \affiliation{E\"otv\"os Lor\'and University, Department of Astronomy, P\'azm\'any P\'eter s\'et\'any 1/A, 1117 Budapest, Hungary}

\author{J. A. Acosta-Pulido} \affiliation{Instituto de Astrof\'\i{}sica de Canarias, Avenida V\'\i{}a L\'actea, 38205 La Laguna, Tenerife, Spain} \affiliation{Departamento de Astrof\'\i{}sica, Universidad de La Laguna,
     38205 La Laguna, Tenerife, Spain} 

\author{M. I. Carnerero} \affiliation{Instituto de Astrof\'\i{}sica de Canarias, Avenida V\'\i{}a L\'actea, 38205 La Laguna, Tenerife, Spain} \affiliation{Departamento de Astrof\'\i{}sica, Universidad de La Laguna,
     38205 La Laguna, Tenerife, Spain} \affiliation{INAF, Osservatorio Astrofisico di Torino, via Osservatorio 20, Pino Torinese, Italy}

\author{ D. Garc\'\i{}a-\'Alvarez} \affiliation{Instituto de Astrof\'\i{}sica de Canarias, Avenida V\'\i{}a L\'actea, 38205 La Laguna, Tenerife, Spain} \affiliation{Departamento de Astrof\'\i{}sica, Universidad de La Laguna,
     38205 La Laguna, Tenerife, Spain} \affiliation{Grantecan S. A., Centro de Astrof\'\i{}sica de La Palma, Cuesta de San
     Jos\'e, E-38712 Bre\~na Baja, La Palma, Spain}

\author{A. Mo\'or} \affiliation{Konkoly Observatory, Research Centre
  for Astronomy and Earth Sciences, Hungarian Academy of Sciences,
  Konkoly-Thege Mikl\'os \'ut 15-17, 1121 Budapest, Hungary}

\author{B. Cseh} \affiliation{Konkoly Observatory, Research Centre
  for Astronomy and Earth Sciences, Hungarian Academy of Sciences,
  Konkoly-Thege Mikl\'os \'ut 15-17, 1121 Budapest, Hungary}

\author{G. Hajdu} \affiliation{Instituto de Astrof\'\i{}sica, Facultad de F\'\i{}sica,
     Pontificia Universidad Cat\'olica de Chile, Av. Vicu\~na Mackenna
     4860, Santiago, Chile} \affiliation{Astronomisches Rechen-Institut, Zentrum f\"ur Astronomie der
		Universit\"at Heidelberg, M\"onchhofstrasse 12-14,
		69120 Heidelberg, Germany} \affiliation{Instituto Milenio de Astrof\'\i{}sica, 
Santiago, Chile}

\author{O. Hanyecz} \affiliation{Konkoly Observatory, Research Centre
  for Astronomy and Earth Sciences, Hungarian Academy of Sciences,
  Konkoly-Thege Mikl\'os \'ut 15-17, 1121 Budapest, Hungary} \affiliation{E\"otv\"os Lor\'and University, Department of Astronomy, P\'azm\'any P\'eter s\'et\'any 1/A, 1117 Budapest, Hungary}

\author{J. Kelemen} \affiliation{Konkoly Observatory, Research Centre
  for Astronomy and Earth Sciences, Hungarian Academy of Sciences,
  Konkoly-Thege Mikl\'os \'ut 15-17, 1121 Budapest, Hungary}

\author{L. Kriskovics} \affiliation{Konkoly Observatory, Research Centre
  for Astronomy and Earth Sciences, Hungarian Academy of Sciences,
  Konkoly-Thege Mikl\'os \'ut 15-17, 1121 Budapest, Hungary}
  
\author{G. Marton} \affiliation{Konkoly Observatory, Research Centre
  for Astronomy and Earth Sciences, Hungarian Academy of Sciences,
  Konkoly-Thege Mikl\'os \'ut 15-17, 1121 Budapest, Hungary}
   
\author{Gy. Mez\H{o}} \affiliation{Konkoly Observatory, Research Centre
  for Astronomy and Earth Sciences, Hungarian Academy of Sciences,
  Konkoly-Thege Mikl\'os \'ut 15-17, 1121 Budapest, Hungary}

\author{L. Moln\'ar} \affiliation{Konkoly Observatory, Research Centre
  for Astronomy and Earth Sciences, Hungarian Academy of Sciences,
  Konkoly-Thege Mikl\'os \'ut 15-17, 1121 Budapest, Hungary}

\author{A. Ordasi} \affiliation{Konkoly Observatory, Research Centre
  for Astronomy and Earth Sciences, Hungarian Academy of Sciences,
  Konkoly-Thege Mikl\'os \'ut 15-17, 1121 Budapest, Hungary}
   
\author{G. Rodr\'\i{}guez-Coira} \affiliation{Instituto de Astrof\'\i{}sica de Canarias, Avenida V\'\i{}a L\'actea, 38205 La Laguna, Tenerife, Spain} \affiliation{Departamento de Astrof\'\i{}sica, Universidad de La Laguna,
     38205 La Laguna, Tenerife, Spain} 

\author{K. S\'arneczky} \affiliation{Konkoly Observatory, Research Centre
  for Astronomy and Earth Sciences, Hungarian Academy of Sciences,
  Konkoly-Thege Mikl\'os \'ut 15-17, 1121 Budapest, Hungary}
     
\author{\'A. S\'odor} \affiliation{Konkoly Observatory, Research Centre
  for Astronomy and Earth Sciences, Hungarian Academy of Sciences,
  Konkoly-Thege Mikl\'os \'ut 15-17, 1121 Budapest, Hungary}
      
\author{R. Szak\'ats} \affiliation{Konkoly Observatory, Research Centre
  for Astronomy and Earth Sciences, Hungarian Academy of Sciences,
  Konkoly-Thege Mikl\'os \'ut 15-17, 1121 Budapest, Hungary}
        
\author{E. Szegedi-Elek} \affiliation{Konkoly Observatory, Research Centre
  for Astronomy and Earth Sciences, Hungarian Academy of Sciences,
  Konkoly-Thege Mikl\'os \'ut 15-17, 1121 Budapest, Hungary}

\author{A. Szing} \affiliation{Baja Observatory, University of Szeged, 6500 Baja, KT: 766}

\author{A. Farkas-Tak\'acs} \affiliation{Konkoly Observatory, Research Centre
  for Astronomy and Earth Sciences, Hungarian Academy of Sciences,
  Konkoly-Thege Mikl\'os \'ut 15-17, 1121 Budapest, Hungary}
       
\author{K. Vida} \affiliation{Konkoly Observatory, Research Centre
  for Astronomy and Earth Sciences, Hungarian Academy of Sciences,
  Konkoly-Thege Mikl\'os \'ut 15-17, 1121 Budapest, Hungary}
      
\author{J. Vink\'o} \affiliation{Konkoly Observatory, Research Centre
  for Astronomy and Earth Sciences, Hungarian Academy of Sciences,
  Konkoly-Thege Mikl\'os \'ut 15-17, 1121 Budapest, Hungary}

\begin{abstract}
V582~Aur is an FU Ori-type young eruptive star in outburst since $\sim$1985. The eruption is currently in a relatively constant plateau phase, with photometric and spectroscopic variability superimposed. Here we will characterize the progenitor of the outbursting object, explore its environment, and analyse the temporal evolution of the eruption. We are particularly interested in the physical origin of the two deep photometric dips, one occurred in 2012, and one is ongoing since 2016. We collected archival photographic plates, and carried out new optical, infrared, and millimeter wave photometric and spectroscopic observations between 2010 and 2017, with high sampling rate during the current minimum. Beside analysing the color changes during fading, we compiled multiepoch spectral energy distributions, and fitted them with a simple accretion disk model. Based on pre-outburst data and a millimeter continuum measurement, we suggest that the progenitor of the V582 Aur outburst is a low-mass T Tauri star with average properties. The mass of an unresolved circumstellar structure, probably a disk, is 0.04\,M$_{\odot}$. The optical and near-infrared spectra demonstrate the presence of hydrogen and metallic lines, show the CO bandhead in absorption, and exhibit a variable H$\alpha$ profile. The color variations strongly indicate that both the $\sim$1 year long brightness dip in 2012, and the current minimum since 2016 are caused by increased extinction along the line of sight. According to our accretion disk models, the reddening changed from $A_V$=4.5 mag to 12.5 mag, while the accretion rate remained practically constant. 
Similarly to the models of the UXor phenomenon of intermediate and low-mass young stars, orbiting disk structures could be responsible for the eclipses.
\end{abstract}


\section{Introduction}
\label{intro}

Young eruptive stars, including FU Orionis (FUor) and EX Lupi (EXor) type objects, form a sub-group of low-mass pre-main sequence stars \citep{herbig77}. Their episodic outbursts correspond to an optical brightening of up to 5 magnitudes, and are explained by a dramatic increase of the accretion rate from the inner circumstellar disk onto the star \citep{hk96,audard2014}. The outbursts may last a few months in EXors, and span decades, or even a century, in the case of FUors. Although the light curve shapes are diverse, they usually start with a sudden initial rise, followed by a longer plateau phase, when the flux level is relatively constant or slowly decaying. In some outbursts the plateau period is interspersed by shorter, sometimes quasi-periodic brightening or fading events. The physical processes governing the initial eruption and the subsequent evolution are still debated \citep[for an overview see][]{audard2014}. 

Multiepoch photometric and spectroscopic observations, in particular the timescale and wavelength dependence of the variability, could provide useful constraints on the underlying physics. The temporary brightenings or dimmings may be especially telltale events of the physical processes acting during the outburst. If the local minima are caused by a temporary drop of the accretion rate, they point to outburst theories which are inherently able to halt and re-start the mass transport process on short timescale \citep{ninan2015}. The fading of a FUor is always interesting because of the possibility to witness, for the first time, the end of a FUor outburst when the system returns to the quiescent state \citep[see e.g.][]{kraus2016}. This return has never been observed yet, as in the monitored cases the fading was always followed by a re-brightening, either completely recovering from the minimum \citep[e.g. V899\,Mon,][]{ninan2015}, or being stabilized at an intermediate flux level \citep[e.g. V346\,Nor,][]{kospal2017}. Apart from non-steady accretion, brightness variations in young eruptive stars may also be linked to changes in the circumstellar extinction  \citep[e.g. V1515~Cyg, V1647~Ori, or PV~Cep, ][]{KHV1515Cyg,acosta2007,kun2011}. These events may reflect a rearrangement of the circumstellar environment, possibly triggered by the outburst, in the form of dust evaporation and/or condensation, or transportation of matter by strong outward winds. Via these processes the eruption of young stars may have a profound effect on the structure of the inner disk, potentially influencing the initial conditions of terrestrial planet formation. 

Our group has been conducting a project to study the variability of young eruptive stars and determine the physical origin of the changes. In an analysis of the light curve of PV~Cep we proposed a structural rearrangement of the inner circumstellar disk \citep{kun2011}. Analysing the long-term photometry of HBC~722, we could interpret the observed multiwavelength flux variations in terms of accretion rate changes in the central accretion disk linked to a varying radius of the outbursting zone \citep{kospal2016}. Recently, we completed an investigation to explain the origin of the deep minimum in the light curve of V346 Nor in 2010 \citep{kospal2017}. We argued that at earlier phases accretion and extinction changes were coupled, but the dramatic drop of flux in 2010 was mostly driven by a temporary halt of accretion only. We also published an analysis of V2492~Cyg, where we demonstrated that the extreme deep minima in the light curve are caused by time dependent dust obscuration \citep{kospal2013}.


As a continuation of our efforts on physical analysis of FUor and EXor light curves, here we present a  study of the multiwavelength light variations of the young eruptive star V582 Aur. The object went into outburst at some time between 1980 and 1985 \citep{samus2009}, and was spectroscopically confirmed to belong to the FUor class by \citet{semkov2013}. It is located close to the Galactic plane, and \citet{kun2017} argued that V582 Aur, together with a group of newly discovered low-mass young stellar objects (YSO), is related to the Aur OB 1 association at a distance of 1.32 kpc from the Sun. Adopting a foreground interstellar extinction of $A_V$=1.53 mag in this direction, \citet{kun2017} derived a bolometric luminosity of 150--320 L$_\odot$ for V582 Aur, a typical value for FU Orionis objects. The flux evolution of the outburst until 2013, including a deep minimum in 2012, was presented by \citet{semkov2013}, while optical and near-infrared photometric and spectroscopic measurements on parts of the subsequent evolution were monitored and published by different authors \citep{oh2015,semkov2017}. 
In this work we will analyse new observations of V582 Aur obtained between 2010 and 2017, supplemented by data extracted from archival photographic plates. 
We will characterize the outbursting object and its progenitor, explore its environment, and document the temporal evolution of the eruption. We will focus on understanding the physical processes causing the two deep photometric dips, one occurred in 2012, and one is ongoing since 2016, and discuss their possible connection to the outburst event.

\section{Observations and data reduction}
\label{sec:obs}

\subsection{Archival pre-outburst measurements}
\label{sec:archdata}

Only very few observations are available for the progenitor and the pre-outburst history of V582 Aur. The archival optical photometric data available in the literature were collected by \citet{semkov2013}. In order to better characterize the eruption and outline the rising part of the outburst light curve, we queried the photograpic plate archive of Konkoly Observatory. We found 18 plates, obtained between 1978 November and 1990 August, that covered the position of V582 Aur. The field of view of all plates was 5$\degr$, but the exposure time (and thus the limiting magnitudes) varied significantly from plate to plate. We digitized the plates and checked whether V582~Aur was visible on them. We detected the source on 5 plates, four $V$-band plates from the same night in 1987 and one white light plate from 1990. The object was invisible on the earlier plates (1978--1986). We performed aperture photometry in the digitized images for V582~Aur and for 44 comparison stars within a 15$'$$\times$15$'$ box centered on the target. For photometric calibration we adopted Johnson $V$ magnitudes from the UCAC4 catalog \citep{zacharias2013}. We applied the same $V$-band calibration on the measurement where no filter was used, too. Using the same calibration, we determined an upper limit for one of the earlier V-band plates. The resulting photometric data are listed in Tab.~\ref{tab:phot}.

\subsection{Optical-infrared photometry}
\label{optirphot}

We performed new optical imaging observations with $BVRI$ filters between 2010 September 19 and 2017 December 7, using the 60/90/180 cm Schmidt and the 1m RCC telescopes at Konkoly Observatory (Hungary), as well as the IAC-80 telescope at the Teide Observatory (Canary Islands, Spain). All telescopes were equipped with standard CCD cameras. For a  detailed description of the telescopes and their instrumentation we refer to \citet{kospal2011}. Near-infrared $JHK_{\rm S}$ images were obtained with a sparser cadence during the same period, using the TCS telescope at the Teide Observatory. At three epochs we also obtained near-infrared images with the LIRIS instrument installed on the 4.2~m William Herschel Telescope at the Observatorio del Roque de Los Muchachos (Canary Islands, Spain). LIRIS data on 2011 October 10 were obtained as part of Project SW2010b06 (PI: \'A. K\'osp\'al), and on 2017 January 5 and February 6 as part of Projects SW2016b18 and SW2016b19 (PI: J. A. Acosta-Pulido). For descriptions of TCS and LIRIS we refer to \citet{acosta2007}. All near-infrared images were taken in a 5-point dither pattern. For both the optical and near-infrared images, standard data reduction and aperture photometry were performed in the same way as in \citet{acosta2007} and \citet{kospal2011}. The typical photometric errors were 2--3\%, although there were some LIRIS measurements where the source entered the nonlinear regime of the detector, and these cases were discarded from the photometry. The resulting optical and near-infrared magnitudes, and their individual uncertainties are listed in Tab.~\ref{tab:phot}. 

\subsection{Optical-infrared spectroscopy}

We obtained low-resolution optical spectra of V582 Aur using the OSIRIS tunable imager and spectrograph \citep{cepa2003} at the 10.4\,m Gran Telescopio Canarias (GTC, Canary Islands, Spain), on 2012 April 16 and 2013 October 18. The observations were part of the service mode  ‘filler’ programs GTC55/12A and GTC3/13A. We used the standard 2$\times$2-binning with a readout speed of 200 kHz.
All spectra were obtained with the R2500R (red) grism,
covering the 5575--7685\,\AA\ wavelength range. Because of the
highly variable seeing we used the 1$\farcs$3 slit oriented at the parallactic angle to minimize losses due to atmospheric dispersion, providing a dispersion of 1.6~\AA/pixel. The resulting wavelength resolution, measured on arc lines, was R=2475. The exposure time was 60\,s in 2012 and 90\,s in 2013. The spectra were reduced and analysed using standard IRAF routines.

Intermediate-resolution spectra of V582 Aur were obtained on 2012 March 31 with the CAFOS instrument on the 2.2-m telescope of the Calar Alto Observatory. For comparison, we also took  spectra of FU~Ori, the prototype of the FUor class. The R-100 grism covered the 5800--9000 \AA\ wavelength range. The spectral resolution of CAFOS, using a 1$\farcs$5 slit, was R=$\lambda$/${\Delta}{\lambda}\approx$ 3500 at $\lambda$ = 6600 \AA. The spectrum of a He-Ne-Rb lamp was regularly observed for wavelength calibration. Broad-band $VR_{\rm C}I_{\rm C}$ photometric images were taken immediately before the spectroscopic exposures for flux calibration. We reduced and analysed the spectra using standard IRAF routines. 

Long-slit near-infrared spectra were taken at four epochs with LIRIS on the WHT, in the framework of the Projects SW2010b06, SW2016b18, and SW2016b19 (see Sect.~\ref{optirphot}). On 2011 September 18 we obtained low resolution spectra in the $ZJ$ and $HK$ bands, with exposure times of 40\,s and 24\,s, respectively. We adopted the 0$\farcs$75 slit width, which yielded a spectral resolution of R=550--700 in the 0.9--2.4$\,\mu$m range. For telluric calibration we observed the B8-type star HIP 25357. On 2017 January 5, 2017 February 6 and 2017 February 8 separate medium resolution spectra in the $J$, $H$, and $K$-bands were obtained with the  1$\farcs$00 slit width, resulting in a spectral resolution of R=2500 in the 1.17--2.41$\,\mu$m range. We observed the A0V star HIP 25593 as telluric calibrator. The measurements were performed with an ABBA nodding pattern. The exposure times were 360-400\,s, 200\,s, and 160\,s in the three bands, respectively. The data reduction was performed in the same way as in \citet{acosta2007}. The
spectra were flux calibrated using $JHK_S$ photometry taken close in time (either by LIRIS or TCS) to the spectral observations. The typical signal-to-noise ratio of the resulting spectra was between 10 and 30.


\subsection{Millimeter observations}
V582 Aur was observed with the Northern Extended Millimeter Array (NOEMA) on 2016 July 31, and on 2016 August 2, 10 and 13 (Project S16AQ, PI: \'A. K\'osp\'al). The target was observed with 7 antennas, providing baseline lengths between 15~m and 175~m. The total on-source correlation time was 9.6 hours. We used the 3 mm receiver centered on 109.0918 GHz, halfway between the $^{13}$CO(1-0) and C$^{18}$O(1-0) lines and each line was measured with a 20 MHz wide correlator with 39 kHz resolution. We also used the wide-band correlator (WideX) to measure the 2.7 mm continuum emission. The single dish half-power beam width (HPBW) at this wavelength is 45$\farcs$8. Bright quasars were used for radio frequency bandpass, phase, and amplitude calibration, and the flux scale was determined using 3C~84, LkH$\alpha$~101, and MWC~349. The weather conditions were mediocre during the observations, with precipitable water vapor between 2 and 15\,mm. The rms phase noise was usually below 60$\degr$ but always below 80$\degr$, and the flux calibration accuracy is estimated to be around 15\%. 

The single dish observations were performed with the IRAM 30m telescope on 2017 March 11--12 (Project 082-16, PI: \'A. K\'osp\'al). We obtained Nyquist-sampled 112$''\times$112$''$ on-the-fly maps using the Eight Mixer Receiver (EMIR) in frequency switching mode in the 110 GHz band with the Versatile Spectrometer Array (VESPA) that provided 20 MHz bandwidth with 19 kHz channel spacing. The Fast Fourier Transform Spectrometer (FTS) was used to cover a wide bandwidth of 4 GHz around the center frequency of 109.0918 GHz with a resolution of 200 kHz to search for additional lines in the spectra. The HPBW at this frequency is 22$''$. The weather conditions were good with precipitable water vapor between 1.5 and 10\,mm.

The data reduction of the single-dish spectra was done with the GILDAS-based CLASS and GREG software packages. We identified lines in the folded spectra, discarded those parts of the spectra where the negative signal from the frequency switching appeared, and subtracted a 2nd or 3rd order polynomial baseline. The interferometric observations were reduced in the standard way with the GILDAS-based CLIC application. For the line spectroscopy we merged the VESPA single-dish and the interferometric measurements in the $uv$-space to correctly recover both the smaller and larger scale emission. First we took the Fourier transform of the single-dish images, then determined the shortest $uv$-distance in the interferometric dataset. A regular grid was made that filled out a circle within this shortest uv-distance, then we took the data points at these locations from the Fourier-transformed single-dish image and merged them with the original interferometric dataset. The resulting synthetic beam was 4$\farcs$2$\times$2$\farcs$9 at PA=16.9$^\circ$. For the continuum only the NOEMA observations were combined, resulting in a synthetic beam of 3$\farcs$0$\times$2$\farcs$7, PA=45.6$^\circ$. After this step, the imaging and cleaning was done in the usual manner.

\section{Results}
\label{sec:res}

\subsection{The environment of V582 Aur}
\label{sec:morp}

\begin{figure}
\centering \includegraphics[width=0.8\columnwidth]{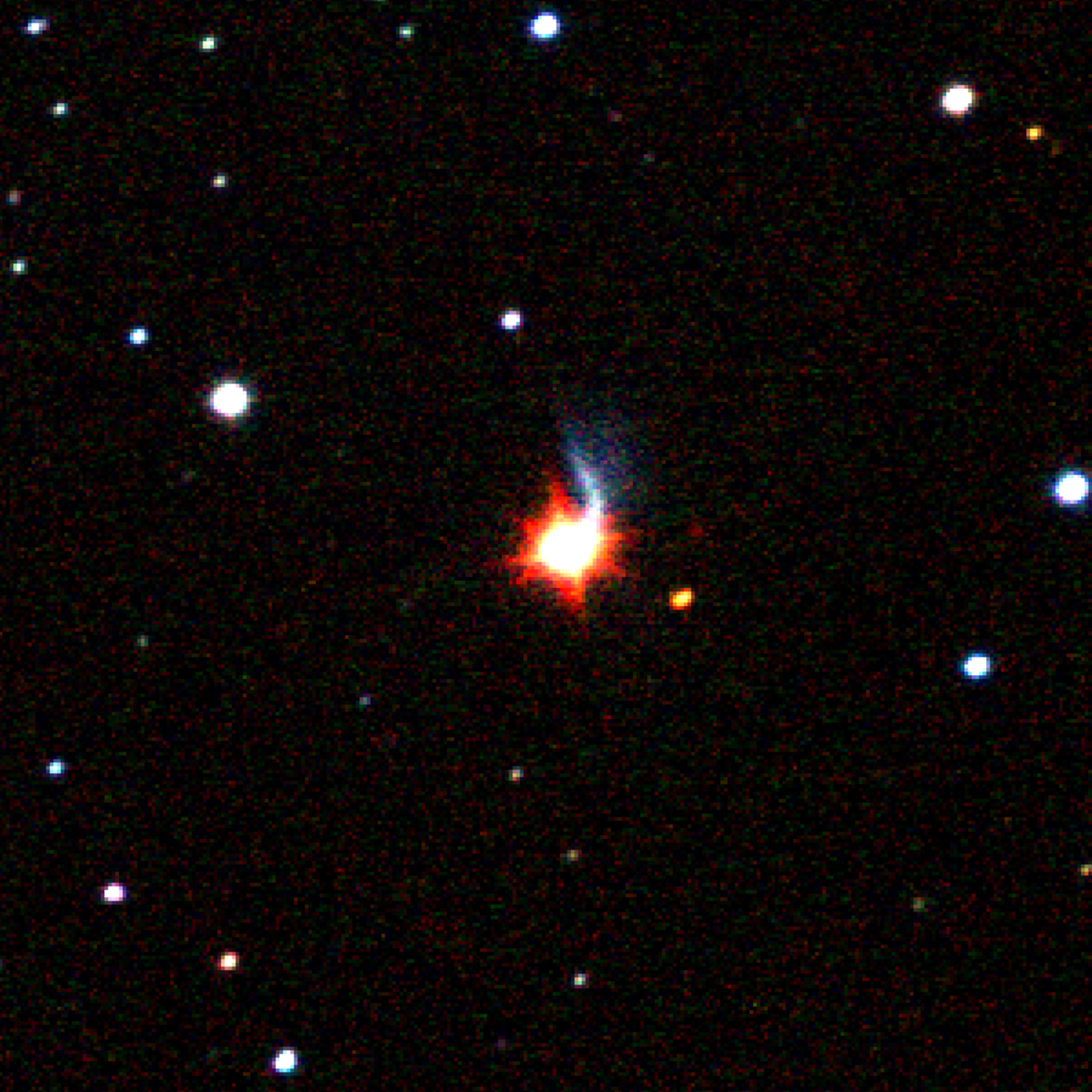}
\caption{Near-infrared color composite image of a 15$'$$\times$15$'$ area centered on V582~Aur, using $J$ as blue, $H_C$ as green, and $K$$_C$ as red. 
The observations were taken with WHT/LIRIS on 2017 February 6. North is up and east is left.}
\label{fig:liriscomp}
\end{figure}

At optical and near-infrared wavelengths V582 Aur is associated with a faint reflection nebulosity (Fig.~\ref{fig:liriscomp}). It has an arc-like filamentary appearance extending to $\approx$7$''$ to the north, corresponding to about 10,000 au at the distance of V582 Aur. The eastern side of the filament is sharp and well defined, while the western side is somewhat more diffuse. Similar structures on comparable spatial scales have been observed around FUors \citep[e.g. Z~CMa,V900~Mon;][]{liu2016,reipurth2012}. Remarkably, the color of this filament is bluer than the source itself, and its brightness and shape are constant over the whole outburst, irrespectively of the deep minima of V582 Aur. In the pre-outburst Palomar Observatory Sky Survey (POSS) blue image from 1954 the filament was not visible. This plate has similar sensitivity and noise values to the later POSS blue plate obtained in 1993, when the filament was clearly seen. Thus, a nebulosity with surface brightness comparable to the 1993 level would have been detected also in 1954. A more detailed analysis of the first POSS image showed that no extended emission was detectable in the direction of the filament above the 1.5$\sigma$ level in 1954.

\begin{figure*}
\centering \includegraphics[angle=90,width=0.8\textwidth ]{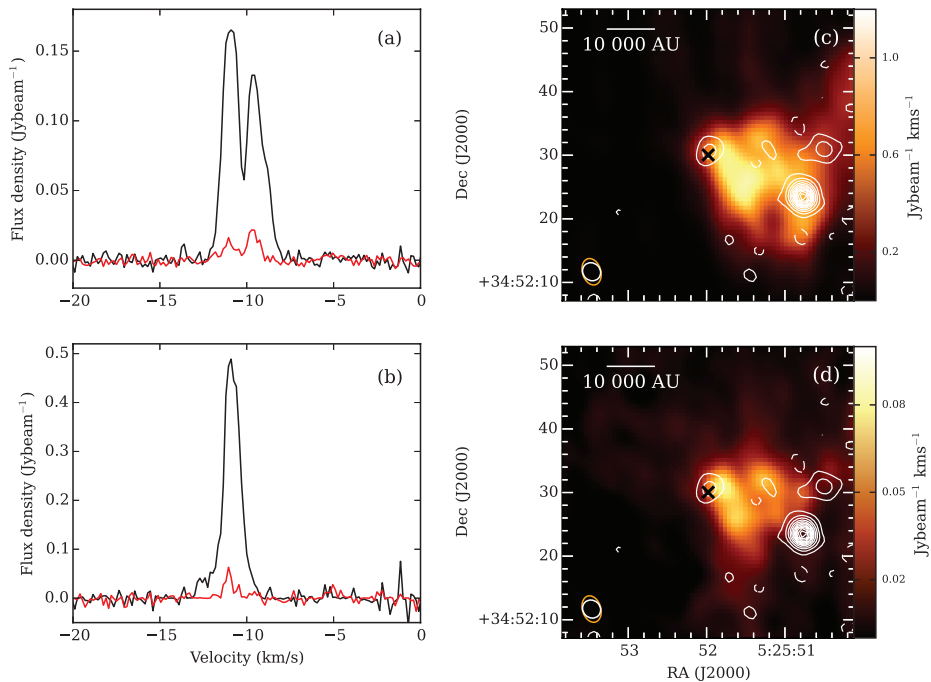}
\caption{IRAM millimeter observations of a region centered on V582 Aur. (a): $^{13}$CO (black) and C$^{18}$O (red) spectra averaged over the 45.8$''$ primary beam. (b) the same spectra of the central synthetic beam. (c) $^{13}$CO (1--0) map integrated between $-12.7$ and $-9.4$ km s$^{-1}$. Contours correspond to the  2.7\,mm continuum emission. The position of the FUor is marked by black 'x' symbol. (d) the same as panel c, but for C$^{18}$O.}
\label{fig:iram}
\end{figure*}

\begin{figure*}
\centering \includegraphics[width=0.8\textwidth]{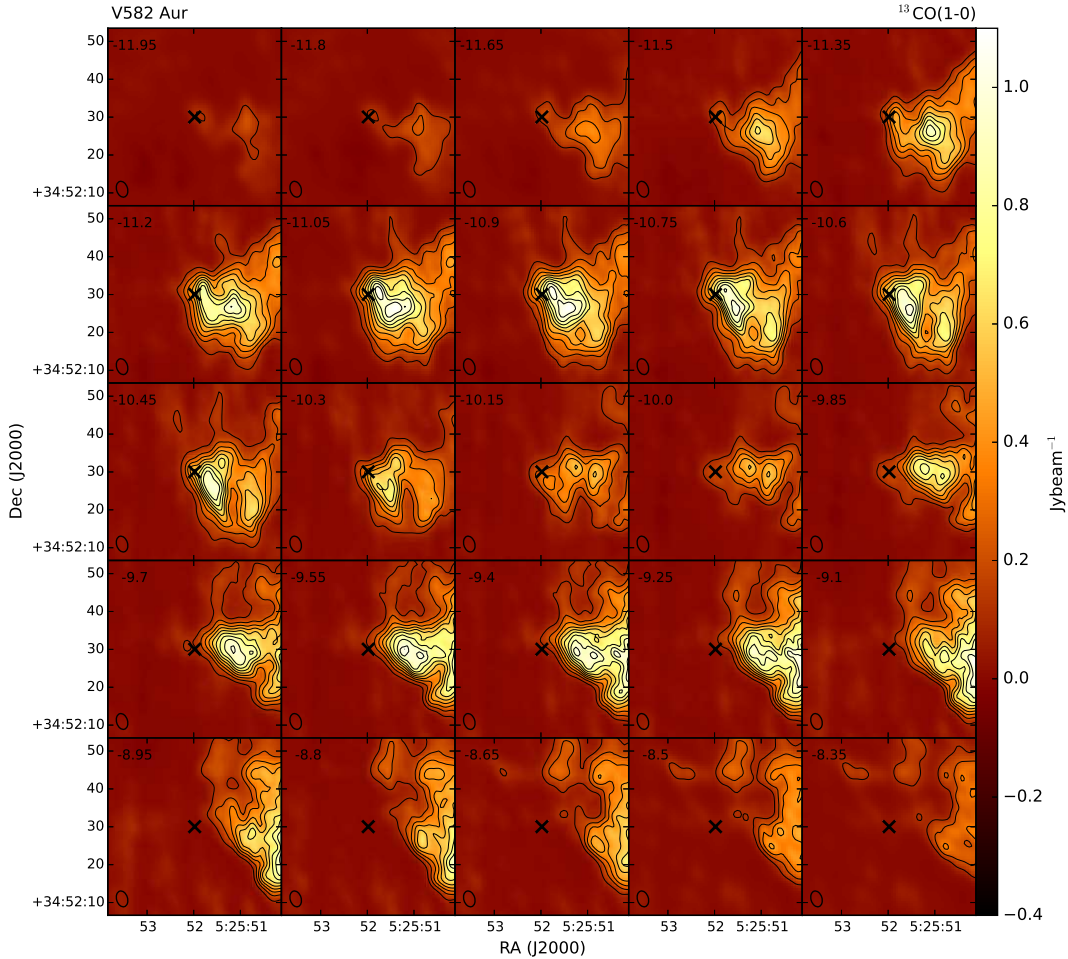}
\caption{Channel maps of the $^{13}$CO(1$-$0) emission around V582\,Aur. The FUor is marked with a cross. The contours are at the 10, 20, \ldots 90\% of the maximum, 1.1\,Jy\,beam$^{-1}$.}
\label{fig:iramch1}
\end{figure*}

\begin{figure*}
\centering \includegraphics[width=0.8\textwidth]{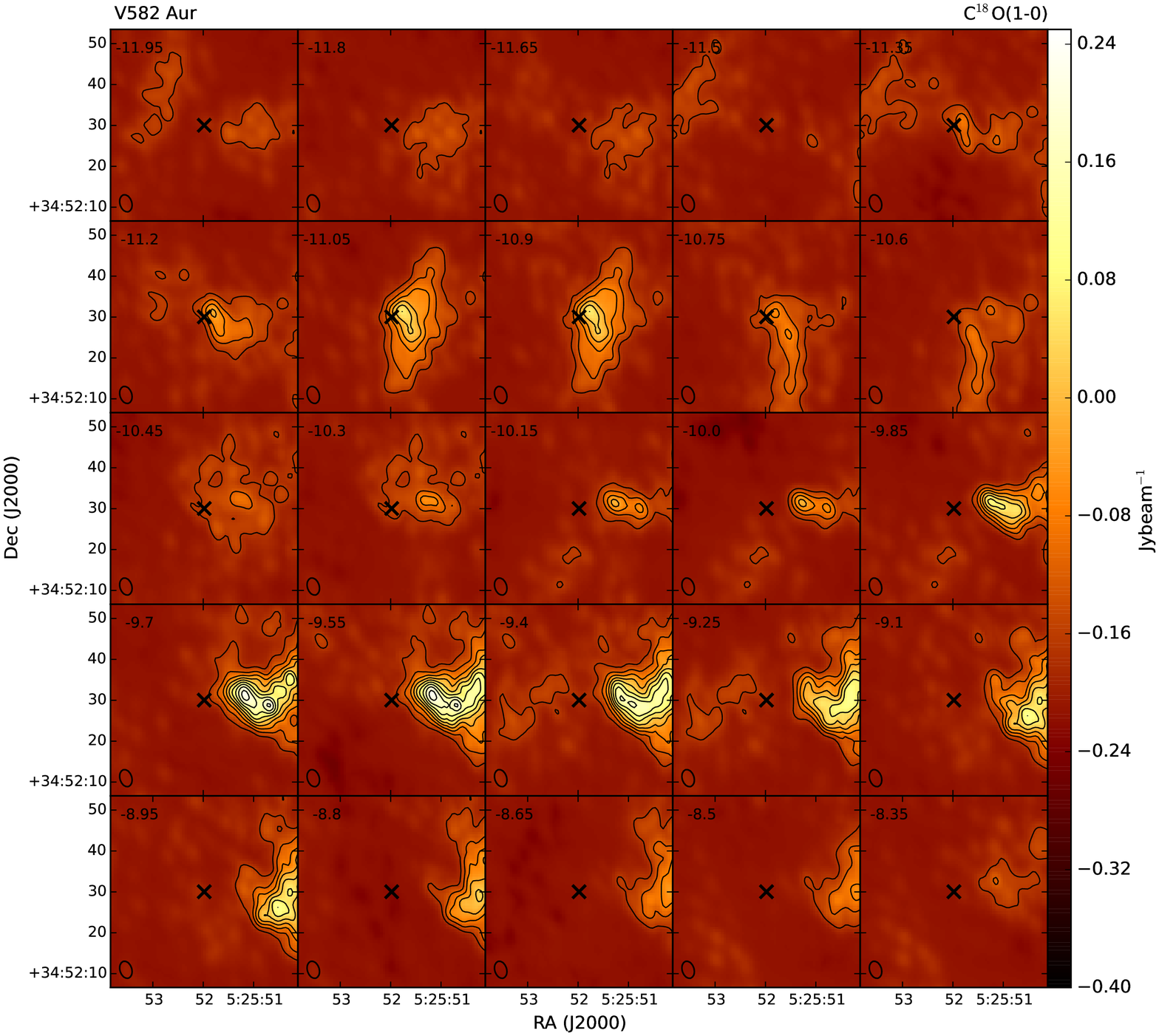}
\caption{Channel maps of the C$^{18}$O(1$-$0) emission around V582\,Aur. The contours are at the 10, 20, \ldots 90\% of the maximum, 0.25\,Jy\,beam$^{-1}$.}
\label{fig:iramch2}
\end{figure*}

Figs.~\ref{fig:iram}, \ref{fig:iramch1}, and \ref{fig:iramch2} present our millimeter maps of the region around V582 Aur. The left panels of Fig.~\ref{fig:iram} show our $^{13}$CO(1--0) and C$^{18}$O(1--0) spectra averaged for the primary beam (upper panel), or measured in the central beam toward the FUor (lower panel). In the upper panel both the $^{13}$CO and C$^{18}$O profiles are similar, exhibiting two separate velocity peaks. In the central beam, towards the FUor, only the more negative component ($-12.7$ $<v_{LSR}<$ $-9.4$\,km s$^{-1}$) is present, as demonstrated in Figs.~\ref{fig:iramch1} and \ref{fig:iramch2}. This velocity range is very similar to the radial velocity of the Aur~OB1 association, $v_{LSR }=-10.5$\,kms$^{-1}$ \citep[$v_{hel}$$=-1.9$\,kms$^{-1}$,][]{melnik2009}, therefore we suggest that this more negative velocity component could be related to V582~Aur. The spatial distribution of this gas component is plotted in the right panels of Fig.~\ref{fig:iram}, revealing diffuse emission as well as a clump situated $\approx$3$''$ (about one beam size) to the west of the FUor. We also overplotted with contours the distribution of the 2.7\,mm dust continuum emission. The continuum data reveal three compact sources, one coinciding with the position of the FUor. A comparison with the synthetic beam shows that the FUor is unresolved in the continuum, and we will assume that the measured emission originates from a circumstellar disk.

The integrated continuum flux for the FUor component is 0.56$\pm$0.12\,mJy. We calculated a total (gas + dust) mass by using Eq.~1 from \citet{ansdell2016}, adopting ${\kappa}_0$=10\,cm$^2$g$^{-1}$ at 1000\,GHz \citep{beckwith1990} with $\beta$=1, and $T$=30\,K, realistic values for a circumstellar disk around a relatively high luminosity source. The gas-to-dust ratio was fixed to 100. The resulting circumstellar mass is 0.04\,$M_{\odot}$. Concerning the gas component, the calculated optical depth values imply that both the $^{13}$CO and C$^{18}$O emission are optically thin \citep[for computational details see][]{feher2017}. The calculated total mass from the central beam (corresponding to  an area of radius$\approx$2800\,au), assuming $T$=30\,K for the temperature in LTE, is 0.054\,$M_{\odot}$ from the $^{13}$CO and 0.028\,$M_{\odot}$ from the C$^{18}$O map. Repeating this calculation for a larger area which includes the CO clump west of the FUor, the total mass within 6$''$ radius (8000\,au) is 0.25\,$M_{\odot}$ from the $^{13}$CO and 0.14\,$M_{\odot}$ from the C$^{18}$O map. For this larger region we assumed a temperature of $T$=20\,K, the average envelope temperature we found in a study of a sample of FUors \citep{feher2017}. Considering all possible uncertainties related to the integration area, the CO abundance ratio, and the gas and dust temperatures, we conclude that our mass results, derived from the molecular line and the dust continuum maps, are in good agreement. The obtained central mass is in the range typical for disks around classical T\,Tauri stars.


The continuum and the CO maps exhibit several additional sources in the neighbourhood of V582 Aur, although the dust and gas clumps do not fully correspond to each other. The continuum map shows two more compact sources, with integrated fluxes of 3.31$\pm$0.11\,mJy (stronger source to the southwest) and 0.73$\pm$0.15\,mJy (source to the west). A detailed survey for young stellar objects in the vicinity of V582 Aur has been recently published by \citet{kun2017}. They identified 68 candidate low-mass stars in an area of 12$'$$\times$12$'$. Only one object from their list falls in the area plotted in Fig.~\ref{fig:iram}. It is mentioned as ``an extended UKIDSS source considered as candidate YSOs'', and is probably associated with the southwest continuum source in our map.

\subsection{The initial brightening and the progenitor}
\label{sec:sed}

\begin{figure}
\centering \includegraphics[width=\columnwidth]{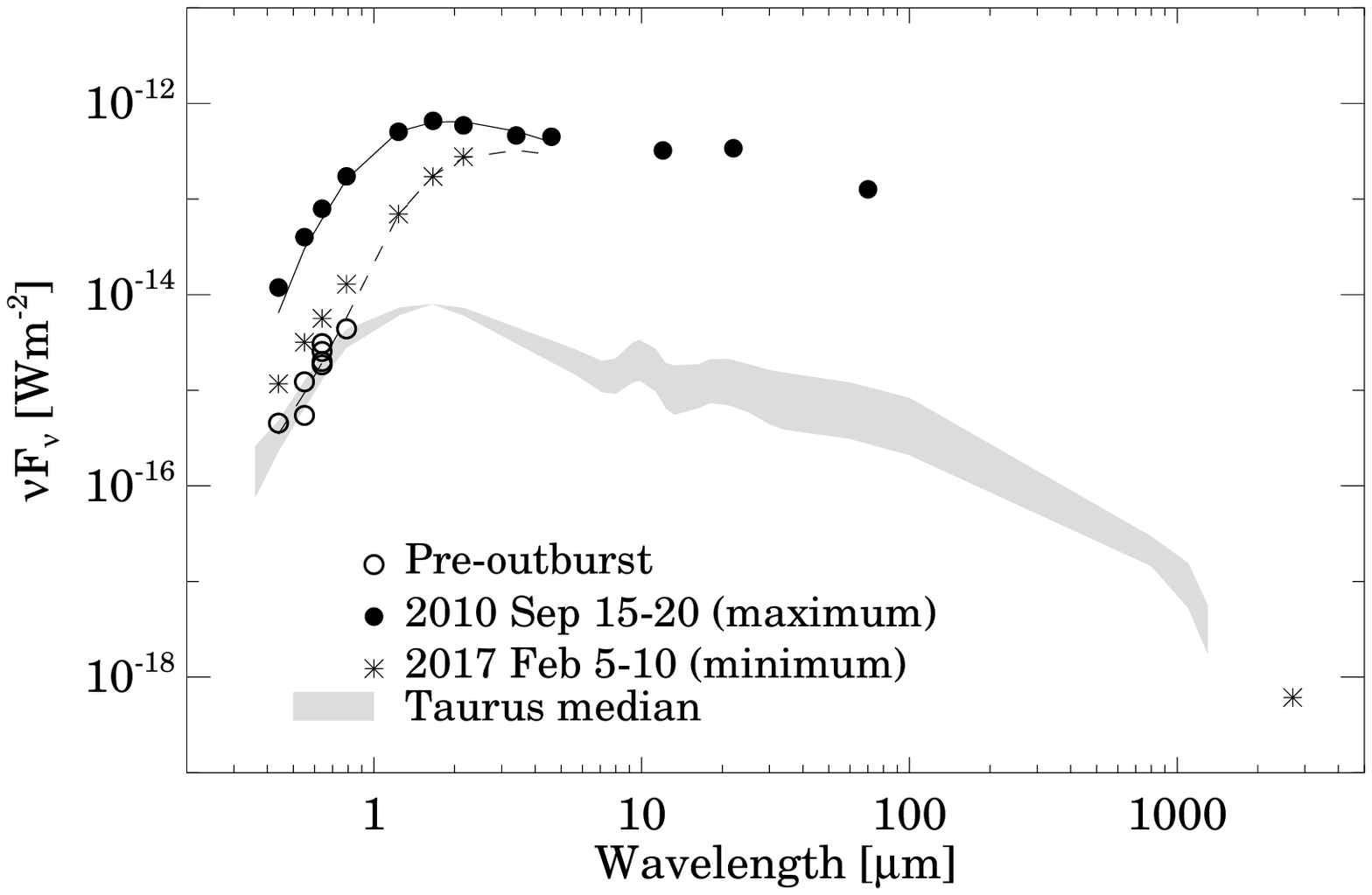}
\caption{Optical-infrared-millimeter spectral energy distributions of V582~Aur at different phases of the eruption. 
Pre-outburst optical data are from \citet{semkov2013}. 
Data for 2010 September, representative of the maximum light of the outburst (Fig.~\ref{fig:light}), and for 2017 February, the minimum of the present dip, are from this work.
The optical and near-infrared data in 2010 are supplemented by WISE observations from 2010 March 8 and 2010 September 15 (Sect.~\ref{sec:light}), and by a Herschel PACS 70\,$\mu$m data point obtained on 2012 Aug 27 \citep{marton2017}. The shaded area marks the Taurus median scaled to the distance of V582 Aur, and reddened by $A_V$=1.53\,mag interstellar extinction (data below 1.25\,$\mu$m and above 34\,$\mu$m are from \citet{dalessio1999}, otherwise from \citet{furlan2006}).}
\label{fig:sed}
\end{figure}

In our investigation of archival photographic plates (Sect.~\ref{sec:archdata}) V582\,Aur was not detected before 1986 February, then became visible on plates obtained in 1987 January. Focusing only on the $V$-band plates, where the source was expected to be brighter, we obtained an upper limit of $V$$>$16.5 mag on 1985 October 21, and a measurement of $V$=13.85\,mag on 1987 January 24, implying that between these two dates the source brightened by ${\Delta}V$$>$2.65\,mag. According to the light curves and photometric table of \citet{semkov2013}, the source was still faint in the $I$-band on 1986 January 17, but was bright in the $B$-band on 1986 December 29, which is consistent with our results. Thus, we can conclude that V582~Aur erupted at some time between 1986 January and December. 

Since V582 Aur has been in outburst for decades, very limited information is available on its pre-outburst state. In Fig.~\ref{fig:sed} we plotted all early photometric points published in \citet{semkov2013}.
The large scatter in those bands where multiple observations are available may be due to  variability, or may be related to the fact that in quiescence V582 Aur was close to the detection limit of the photograpic plates. In the figure we overplotted the ``Taurus median'', a representative spectral energy distribution (SED) of classical T Tauri stars in the Taurus star forming region. Before plotting, we scaled the median fluxes to 1.32 kpc, the distance of V582 Aur, and reddened them by $A_V$=1.53\,mag, the interstellar extinction toward V582~Aur according to \citet{kun2017}. The pre-outburst observations, within their dispersion, match the Taurus median both in the absolute flux and in the shape of the SED. We also overplotted our 2.7 mm continuum flux measurement (Sect.~\ref{sec:morp}), which seems to be consistent with an extrapolation of the Taurus median (if we extrapolate it from the 100--1300\,$\mu$m slope and ignore the sharp turndown at 1300\,$\mu$m). V582~Aur appears as a point source in our dust continuum measurement, and the agreement with the Taurus median suggests that the large part of the matter around V582\,Aur is concentrated in a circumstellar disk. Thus, we will consider the total (gas + dust) mass of 0.028--0.054\,M$_{\odot}$, derived in Sect.~\ref{sec:morp}, as the mass of a circumstellar disk. Since to our knowledge no other information is available on the quiescent period, on the basis of the above results we propose that the most likely progenitor of the V582 Aur FUor eruption was a very typical late-type T Tauri star.

\subsection{Light curves and the evolution of the outburst}
\label{sec:light}

\begin{figure*}
\centering \includegraphics[height=\textwidth,angle=90]{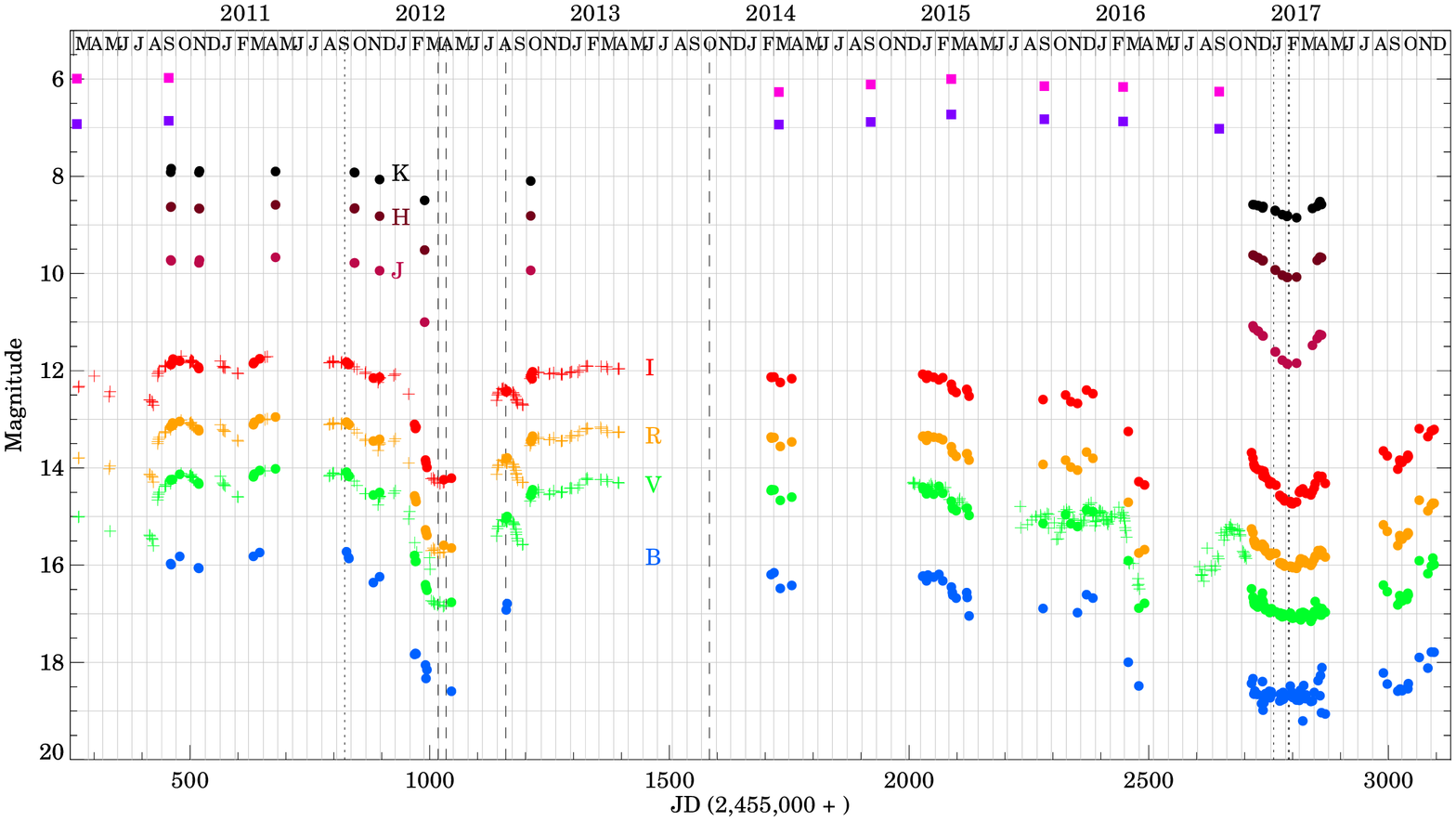}
\caption{Light curves of V582~Aur. Filled dots are from this work,
  plus signs are from \citet{semkov2013}, and from the ASAS-SN sky survey
  \citep{shappee2014,Kochanek2017}. Mid-IR data points are from WISE
  (filled squares).
  Tick marks on the top indicate the first
  day of each month. Vertical dotted (dashed) lines mark the epoch when our
  near-infrared (optical) spectra were taken.}
\label{fig:light}
\end{figure*}

Figure~\ref{fig:light} displays the multiband light curves of V582~Aur between 2010 and 2017, compiled from our observations and from literature data. At mid-infrared wavelengths we plotted photometry obtained by the WISE space telescope \citep{wright2010} in its cryogenic mission, and in the NEOWISE and NEOWISE-Reactivated periods. For each WISE epoch, we downloaded all time resolved observations from the AllWISE Multiepoch Photometry Table and from the 
NEOWISE-R Single Exposure Source Table in the W1 (3.4\,$\mu$m) and W2 (4.6\,$\mu$m) photometric bands, and computed their average after removing outlier data points. In the errors, we added in quadrature 2.4\% and 2.8\% as the uncertainty of the absolute calibration in the W1 and W2 bands, respectively (Sect. 4.4 of the WISE Explanatory Supplement). 

According to the long term historical light curve presented in \citet{semkov2013}, and also to our data (Sect.~\ref{sec:sed}), the initial flux rise of V582 Aur occurred in 1986. The amplitude of the brightening was 4--5 mag in the V-band. We have very limited information on the outburst between 1986 and 2010. The detailed multiwavelength light curve in Fig.~\ref{fig:light} reveals a relatively constant period between 2010 and mid-2012. Afterward, the system entered a deep minimum, from which it recovered by 2013 October. The minimum brightness was only 1.0--1.5 mag above the pre-outburst level \citep{semkov2013}. Another relatively constant period lasted until late 2014, when a shallower dip, again about one year long, started. Before the complete recovering of the system, began the most spectacular, and is still ongoing feature of the light curve. In late February - early March 2016 the brightness dropped to the level of the 2012 minimum within a month. During the summer period the object became brighter by V$\sim$0.8 mag with an increasing trend. However, after a local peak in October 2016, a new extended minimum developed, which reached its deepest point in February 2017. Since then the FUor was brightening until 2017 April. Our latest observations from 2017 August-December suggest that the rising trend may be continued, though local fluctuations and short timescale brightness drops might also occur.

The shapes of the light curves are very similar, with no apparent time delay between them. The amplitude of the variability, however, depends on the wavelength. During the current minimum, this dependence shows an unexpected pattern: the deepest minimum is seen in the I- and J-bands, while at both shorter and longer wavelengths the amplitude of the fading is smaller. The mid-infrared light curve is sporadic to draw firm conclusions on the variability of the thermal emission component probably originating from the innermost regions of the circumstellar matter, although in general it seems to follow the pattern shown by the $V$-band light curve. There are also hints for short timescale variability in the optical data. As an example, in 2012 September, at the end of the 2012 minimum, the system has almost returned to the maximum state when a temporary shallower minimum occurred in all observed bands. 

Both the optical and infrared images display an asymmetric nebulosity around the star. Our color composite image in Fig.~\ref{fig:liriscomp} follows a similar morphology than the optical image of \citet{semkov2013}, including the filament to the north (Sect.~\ref{sec:morp}). It is interesting that the brightness of this filament, unlike that of the star, did not change considerably between the LIRIS measurements obtained in 2011 (in maximum) and in 2017 (in minimum). At a distance of 1.32 kpc, the $\approx$7$''$ length of the filament would correspond to 0.15 light year in the plane of the sky, but its physical extent may be larger depending on the inclination. If the present brightness minimum of V582~Aur, which began in late February, 2016, was caused by a drop of the luminosity of the central source, a corresponding fading of the illuminated filament is expected with a time-lag dictated by the light travel time. The 10 months time difference between the start of the present minimum and our 2017 January LIRIS image should have produced an observable decrease of the surface brightness of the filament, unless its inclination is nearly parallel (within 10 degrees) to the line-of-sight. Thus, the invariable brightness of the filament suggests that the luminosity of the central source did not change during the current 2016--17 minimum.

\subsection{Optical spectroscopy}
\label{sec:optspec}

\begin{figure}
\centering \includegraphics[height=\columnwidth,angle=90]{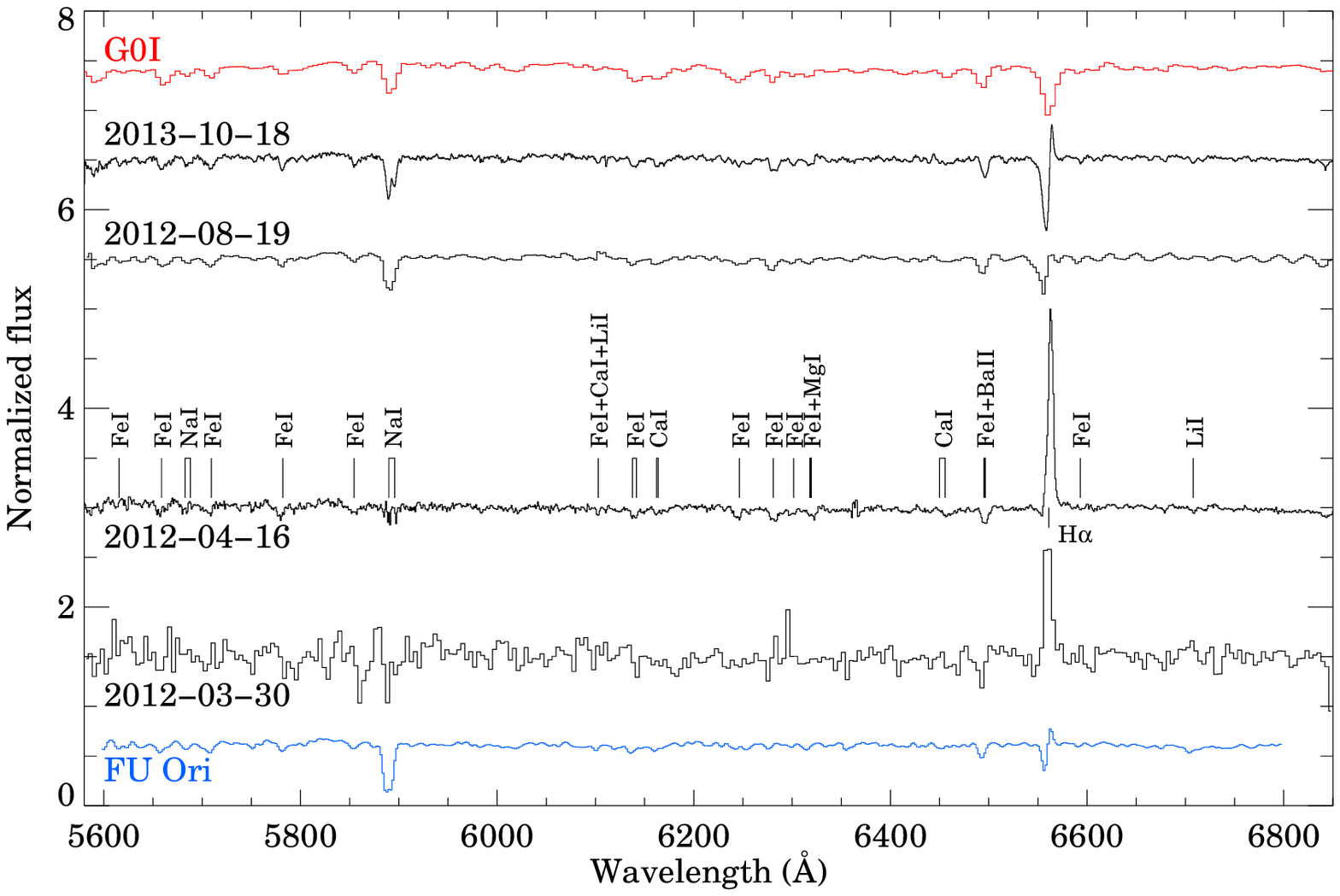}
\caption{Low-resolution optical spectra of V582~Aur at four different epochs, compared to a G0I-type stellar spectrum from \citet{pickles1998} and the spectrum of FU~Ori. 
}
\label{fig:spectra}
\end{figure}

We obtained four optical spectra close in time to the brightness minimum of the source in 2012. Two measurements were taken during the minimum, one in a re-brightening period, and one after the dip, already at maximum light. Fig.~\ref{fig:spectra} shows our spectra. For comparison we overplotted a G0I supergiant spectrum \citep{pickles1998} and our spectrum of FU\,Ori, the prototype of the FUor class. Many neutral metallic lines (Fe, Na, Ca, Mg) are visible, always in absorption, as typical for FUors \citep{hk96}. The 6708~\AA\ lithium line, typical for YSOs, is present. These spectral features seem to be present at all epochs, with similar strength. They also agree well with the spectrum of the G0I star, specifying an effective spectral type at brightness maximum. The signal-to-noise ratio of the FU Ori spectrum is not very high, nevertheless the detected features are also present in our V582~Aur spectra. 

The H$\alpha$ line shows a clear P~Cygni profile, also characteristic of FUors. The P~Cygni profile is less visible in the low S/N spectrum obtained with CAFOS at high airmass on 2012 March 30, near the bottom of the light curve. The ratio of the emission and absorption components is strongly variable: the emission component dominates in the spectra obtained around the photometric minimum in 2012, whereas a deep and wide blue-shifted absorption, accompanied by a narrow emission component appear in the spectra obtained during the bright state on 2012 August 19 and 2013 October 18. In order to check whether the variation originated from changing wind strength, we compared the fluxes of the H$\alpha$ emission component in the low- and high-state spectra. We measured an equivalent width (EW) of $11.8\pm0.5$~\AA\ in the OSIRIS spectrum observed on 2012 April 16, and $0.96\pm0.05$~\AA\ in the spectrum observed on 2013 October 18 with the same instrument. The underlying continuum flux, estimated from nearly simultaneous R magnitudes (15.74 and 13.2 mag, respectively) increased by a factor of some 10.4 between the two epochs. Therefore, the EWs of the H$\alpha$ emission components and the underlying continuum changed by a similar factor, suggesting that the actual line fluxes stayed approximately constant. This implies, that the wind component that gives rise to the emission component remained nearly invariable.

\subsection{Near-IR spectroscopy}

\begin{figure*}
\centering \includegraphics[width=\textwidth,angle=0]{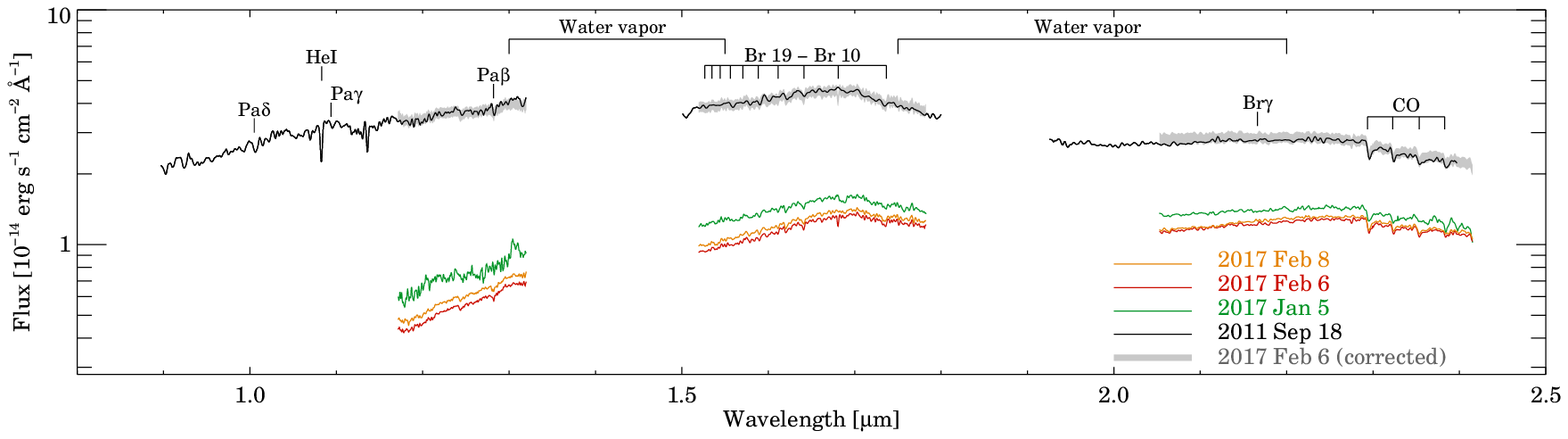}
\caption{Near-infrared spectra of V582~Aur obtained with WHT/LIRIS. The gray strip represents our 2017 February 6 spectrum, dereddened by 7.3 mag.}
\label{fig:liris}
\end{figure*}

The first LIRIS spectrum from 2011 September 18 represents the very end of a relatively constant bright period of the lightcurve, just preceding the deep 2012 minimum. The new, high-resolution spectra from 2017 correspond to the rapid fading (2017 January 5) or to the minimum epoch (2017 February 6 and 8). In Fig.~\ref{fig:liris} the spectral shape of the 2011 Sep 18 spectrum increases from the J to the middle of the H-band, with a discontinuity at the gap between the two bands. The shallow depression in the H-band part of the spectrum is probably related to water vapor absorption. Around 1.7\,$\mu$m there is a peak, with a turn down at longer wavelengths towards the K-band. Another shallow dip appears between the H and the K bands, which is also attributed to water vapor \citep{greene1996, aspin2009}. The same spectral characteristics can be observed also in 2017, apart from a different general slope. Weak atomic lines can be seen in absorption in the spectra, in particular the hydrogen Paschen and Brackett series. The Br$\gamma$ is marginally visible in the 2017 February 6 spectrum, but almost undetectable two days later. A large number of metallic lines (NaI, MgI, SiI) appear in the spectra, although at low significance level. A strong absorption of HeI at 1.083\,$\mu$m is also evident. The CO bandhead is well detected in absorption. The bandheads seem to be associated with a weak blueshifted emission component, which was somewhat stronger in 2011 and weaker in 2017. 

\section{Discussion}
\label{sec:dis}

\subsection{Color changes during the brightness minima}
\label{sec:colors}

\begin{figure}
\centering \includegraphics[width=0.75\columnwidth]{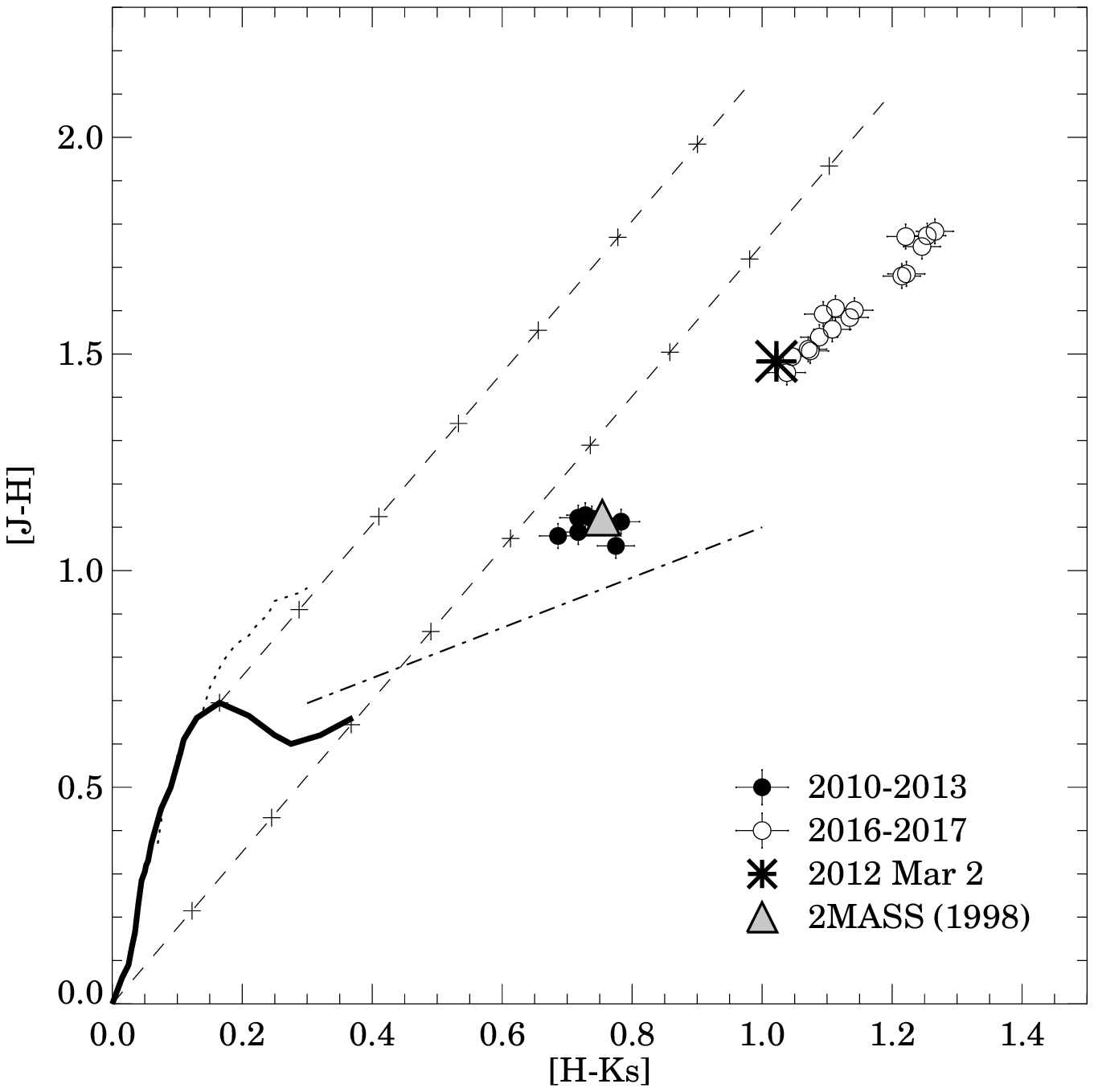}
\caption{Near-infrared color-color diagram of V582 Aur. The main sequence, the giant branch \citep{koornneef1983} and the T Tauri locus \citep{meyer1997} are marked by solid, dotted and dash-dotted lines, respectively. The dashed lines delineate the direction of the reddening path, marked at A$_V$=2, 4, 6\ldots \,mag.}
\label{fig:tcd}
\end{figure}

In order to explore the possible physical origin of the two deep brightness minima in 2012 and 2016-17, we analysed the infrared and optical colors during the fading events. Fig.~\ref{fig:tcd} presents the [J--H] vs. [H--K$_s$] color-color diagram of all near-infrared data points from Fig.~\ref{fig:light}. Overplotted is the path of interstellar extinction with $R_V$=3.1 from \citet{cardelli1989}. The measurements obtained in 2016--17 closely follow the reddening path over the whole dip. The 2012 minimum was not well sampled at near-infrared wavelengths, only one data point was obtained at lower flux level. The location of this point, however, is also consistent with reddening, thus Fig.~\ref{fig:tcd} implies that the main physical mechanism behind both flux declines was variable extinction, most likely by circumstellar dust.

This conclusion is strongly supported by the comparison of the shapes and absolute levels of our multiepoch near-infrared spectra presented in Fig.~\ref{fig:liris}. Selecting the 2017 February 6 spectrum, which exhibits the lowest absolute flux level in our sample, and correcting it for extinction with A$_V$= 0.3, 1.3,and 7.3 magnitudes, we can precisely match  the brighter spectra on 2017 February 8, 2017 January 5, and 2011 September 18, respectively (see Fig.~\ref{fig:liris}). While the agreement in absolute flux levels is less surprising, since the spectra were calibrated by near-infrared photometric points which were shown to follow the reddening path in Fig.~\ref{fig:tcd}, the match in the spectral shape  over the extended wavelength range of 1.17--2.41\,$\mu$m gives a strong support to our hypothesis that the physical mechanism is time-dependent extinction. 

\begin{figure}
\centering \includegraphics[width=\columnwidth]{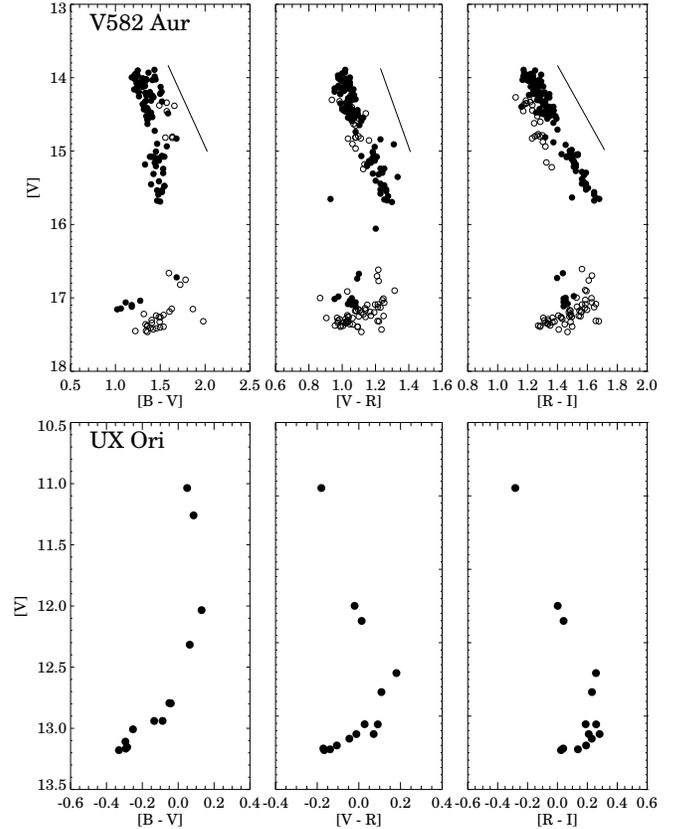}
\caption{Optical color-magnitude diagrams. Upper panels: V582 Aur, compiled from our photometry in Tab.~\ref{tab:phot}. Filled symbols correspond to epoch before 2013, open circles mark observations from 2015--17. The lines correspond to an extinction change of A$_V$=1\,mag. Lower panels: UX~Ori, compiled from our monitoring observations carried out in 2009 October--November (Szak\'ats et al. in prep.)}
\label{fig:cmd}
\end{figure}

Figure~\ref{fig:cmd} summarizes the flux and color variations at optical wavelengths. The upper panels display three kinds of optical color-magnitude diagrams, based on our observations from Tab.~\ref{tab:phot}. While the distribution of data points seems to follow the interstellar extinction during the bright states of the system, it deviates from the linear trend below a magnitude threshold, which is about V=14.5 in the [V] vs. [B-V] diagram and V=15.5 in the other two plots. The sense of the deviation is that when the system is fading in the V-band, its colors turn back and become bluer than what is expected from reddening. For comparison, in the lower panels of Fig.~\ref{fig:cmd} we plotted our optical observations of UX Ori, the prototype UXor. The distribution of points is qualitatively similar to our results on V582~Aur. 
The physical reason invoked to explain the blueing effect in UXors is that the observed optical flux has a contribution from stellar light scattered toward us from circumstellar dust particles. When the star is being eclipsed to a large part by a dust cloud in the line of sight, the observed flux will be dominated by the unobscured scattered component, which is inherently blue \citep{bibothe1990,natta2000}. The similarity of the color evolution in the UXor phenomenon and in the case of V582 Aur suggests a similar geometry: orbiting dust clumps obscure the star, while more distant circumstellar regions, which scatter the stellar light toward us, remain unobscured. 

The combined effect of extinction and blueing provides an explanation for the wavelength dependence of the variability amplitudes in Fig.~\ref{fig:light}. The figure shows that the amplitude at infrared wavelengths increases from the $K_{\rm S}$ to the $J$ band, consistently with the shape of the extinction curve, and with the systematic change of the shape of the infrared spectra at different epochs in Fig.~\ref{fig:liris}. In the $J$ and $I$ bands the amplitudes are comparable, but toward shorter wavelengths the amplitude monotonically decreases, which is not compatible with extinction. It can, however, be well explained by the increasing role of scattering, which is most dominant at the shortest wavelengths causing the blueing effect (Fig.~\ref{fig:cmd}).


The multiwavelength light curves in Fig.~\ref{fig:light} provide constraints on the location and characteristic size of the eclipsing dust cloud. If the mid-infrared emission follows the optical fading, albeit with a lower amplitude in accordance with the interstellar reddening law, then the obscuring dust structure would cover the whole inner part of the circumstellar disk or envelope, where the mid-infrared emission originates. In order to test this possibility, we interpolated the V-band light curve in Fig.~\ref{fig:light} for the epochs of the WISE data points (when optical data existed within 10 days), and analysed the optical--mid-infrared correlation. We found the relationships [W1]$\sim$0.20$\pm$0.11$\times$[$V$], and [W2]$\sim$0.14$\pm$0.12$\times$[$V$], where [W1] and [W2] are the WISE magnitudes. The slopes of the magnitude changes at the optical and mid-infrared wavelengths are significant only at the 1.1--2\,$\sigma$ level. They nevertheless suggest positive correlations, with slopes somewhat higher than the ones expected from the interstellar extinction curve \citep[0.056 and 0.035, respectively,][]{cardelli1989}. Thus, our results hint that the eclipsing dust cloud obscures an area that includes the region where the 3.4--4.6~$\mu$m emission is originated. The higher than expected mid-infrared variability amplitude with respect to the optical one, if confirmed, may be related to the blueing effect of the optical flux, which causes an underestimation of the $V$-band variability amplitude.

\subsection{Accretion disk modeling}
\label{sec:accdisk}

\begin{figure}
\centering \includegraphics[width=\columnwidth]{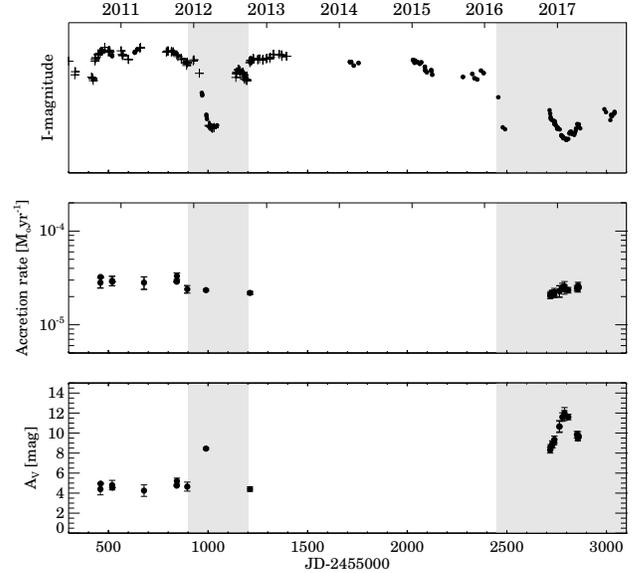}
\caption{Time evolution of the accretion rate (middle panel) and the extinction (lower panel) derived from the accretion disk model fit to the near-infrared spectral energy distribution (Sect.~\ref{sec:accdisk}). The upper panel shows the $I$-band light curve. The shaded areas mark the intervals of brightness minima.}
\label{fig:accdisk}
\end{figure}

A more quantitative approach to separate the physical effects of changing extinction and accretion rate is fitting the SED in each epoch by an accretion disk model. We adopted a steady optically thick and geometrically thin viscous accretion disk, with a radially constant mass-accretion rate (see Eq.~ 1 in \citet{kospal2016}). Similar disk models could reproduce the SEDs of other FUors in the literature \citep{hk96,zhu2007}, and we also used it successfully for HBC~722  \citep{kospal2016} and  for V346~Nor \citep{kospal2017}. In the model the disk SED was calculated via integrating the blackbody emission of concentric annuli between the stellar radius and $R_{out}$.  We fixed the outer radius of the accretion disk to $R_{out}$ = 2 au (the exact value has no noticeable effect on the results), thus we had only two free parameters: the product of the stellar mass and the accretion rate M\.M, and the extinction $A_V$. Since the object is probably a typical low-mass T Tauri star (Sect.\,\ref{sec:sed}), we fixed the stellar mass and radius to 1 M$_{\odot}$ and 3.0 $R_{\odot}$. Because the inclination of the V582 Aur disk is unknown, we adopted 45$\degr$. Finally, we added to the accretion disk model fluxes the Taurus median fluxes, scaled to the distance of V582~Aur, to account for the contribution of the quiescent object (see Fig.~\ref{fig:sed}). The resulting synthetic fluxes were reddened using a grid of A$_V$ values and the standard extinction law from \citet{cardelli1989} with $R_V$ = 3.1.
The fitting procedure was performed with ${\chi}^2$ minimization. 

Since the optical data may be contaminated by scattered light, we fitted only the $JHK_{\rm S}$ data points. Our models reproduced also the optical part of the SED reasonably well, especially in the high flux epochs outside the two deep minima. During the two dips our model systematically underestimated the measured optical fluxes, which we attribute to the presence of a scattered light component in the measurements not accounted for in the model (Fig.~\ref{fig:sed}). In Fig.~\ref{fig:accdisk} we plotted the resulting \.M and $A_V$ values as a function of time (the epochs of the two deep minima are marked by gray stripes). The data outline clear trends. We found that the accretion rate was constant within 15\,\% over the whole period. The average \.M$\sim$2.5$\times10^{-5}$\,$M_{\odot}$yr$^{-1}$ is a rather typical value for FU Orionis objects \citep{hk96,audard2014}. The extinction toward the source is also relatively constant, with $A_V\approx$4.5 mag, apart from the periods of the minima. The first dip in 2012 was not well sampled in the near-infrared domain, we have only one epoch. However, the measured extinction value increased to 8 mag, then returned to 4.5 mag after the minimum ended. During the ongoing brightness minimum of V582 Aur in 2016--17, we covered the flux evolution also at near-infrared wavelengths. A comparison of Fig.~\ref{fig:light} and Fig.~\ref{fig:accdisk} shows that the lower was the brightness of the source the higher was the extinction toward the source while the accretion rate was unchanged. The peak extinction value, measured on 2017 February 7, was 12.5 mag. 

While the exact values of both the accretion rate and the extinction may slightly depend on the assumptions in our disk model, the general trend described above turned out to be very robust. Note that the extinction difference between the states outside the minima and at the bottom of the dip in 2017 February is consistent with the ${\Delta}A_V\approx$7.3\,mag we extracted from the LIRIS spectroscopy in Sect.~\ref{sec:colors}. Thus, according to the conclusions of the previous subsection, we claim that the two deep minima of V582 Aur were caused by an increase of the line-of-sight extinction by 7--8\,mag, while the accretion rate onto the star remained unchanged.


\subsection{Structural changes related to the outburst?}

Our results imply that one or several dust clumps exist in the circumstellar environment of V582~Aur, and these clumps pass the line-of-sight of the star. The documented temporal baseline of the photometric data is not enough to decide whether the two observed minima in 2012 and in 2016--17 were caused by the same orbiting dust structure or by two individual clumps. The two eclipses are rather similar in amplitude, but the current one lasts longer and no repeated patterns can be seen in the light curves during the two minima. In the following we will perform an analysis assuming that the two fadings were caused by the same orbiting dust clump, thus the events are periodic. With this hypothesis, the characteristic distance of the clump from the central star can be estimated by assuming a circular orbit around a 1\,M$_{\odot}$ star, and adopting a period of $\sim$4.75\,yr. This simple calculation gives an orbital radius of 2.8\,au, which is a lower limit if the star is less massive than the Sun. The derived 2.8\,au is larger than the dust sublimation radius, defined as the radial location where the temperature is 1500\,K. For the sublimation radius we estimated 0.07\,au in the pre-outburst phase, and 0.8--1.2\,au in eruption, depending on the actual outburst luminosity within the range proposed by \citet{kun2017}. Thus the existence of silicate particles at the distance calculated from the orbital period is possible even in outburst. 

The comparison of the period of $\sim$4.75\,yr with a characteristic length of the eclipses of $\sim$1 year implies that the clump is spread over about one fifth of the orbit,  $\sim$3.5\,au. When centered on the line-of-sight, this clump could cover the inner disk within 1.75\,au radius. The dust temperature at this radius is 300\,K in quiescence, and 1050--1250\,K in outburst, depending on the luminosity. These results suggest that the area where the dominant part of the 3.4\,$\mu$m and 4.6\,$\mu$m emission in Fig.~\ref{fig:light} comes from is obscured, thus at mid-infrared wavelengths a light curve is expected to be similar to the optical one, but with lower amplitude. Although our WISE light curve is not well sampled, in Sect.~\ref{sec:colors} we argued that there is a weak observational hint for a correlation with the optical variability, lending support to the estimated orbital distance of the clump. The constancy of the scattered light is also consistent with the proposed clump location. The scattered component may originate from all radial locations from the disk, and even further from the extended circumstellar environment including the 7$''$ long filamentary structure to the north (Sect.~\ref{sec:morp}). Thus, the estimated 2.8\,au radial distance of the orbiting clump from the star would imply that a large part of the scattered light would remain unobscured, and its contribution to our spatially unresolved photometric observations may lead to the observed UXor-like blueing effect in Fig.~\ref{fig:cmd}.  

The column density of the clump can be deduced from the 7.3\,magnitude change of the extinction from maximum to minimum (Fig.~\ref{fig:accdisk}). With the relation between optical extinction and hydrogen column density of \citet{guver2009}, A$_V$=7.3\,mag corresponds to 0.026\,gcm$^{-2}$. The total mass of a dust clump whose length is 3.5\,au, and whose height is arbitrarily taken to 1\,au, is about 
1.2$\times$10$^{-8}$\,$M_{\odot}$ or 0.004\,$M_{\oplus}$. Less edge-on line-of-sight would correspond to a larger vertical extent of the clump, thus to a somewhat higher mass. The estimated mass is negligible compared to the matter already accreted onto the star during the FUor outburst, and had to be stored in the inner disk before the eruption. Adopting a representative 2.5$\times$10$^{-5}$\,M$_{\odot}$ yr$^{-1}$ from Fig.~\ref{fig:accdisk}, and 30 years for the outburst, we derived 7.5$\times$10$^{-4}$\,M$_{\odot}$ for the already accreted mass, four orders of magnitude higher than the estimated mass of the eclipsing dust clump. Thus the clump can be considered as a small fluctuation within the inner disk structure. If the obscuring dust clump is not an orbiting but an infalling structure arriving from the outer disk, then its mass would correspond to $\sim$2 million of Hale-Bopp comets, thus it would be a very massive supercomet. Infalling fragments in the protoplanetary disk are predicted in the burst mode models of FUor outbursts \citep{vorobyov2010}. However, those fragments range from several Jovian masses to low-mass brown dwarfs, being much more massive than the dust clump in V582\,Aur.

The observed behavior of V582\,Aur, together with the simple analyses above, seem to suggest that the observed variability is not necessarily connected to the FUor outburst. Similar year-long eclipses were detected in other low-mass, non-erupting young stars, too. AA\,Tau, the prototype of dippers, entered a deep fading event in 2011 and is still in the low state. The event is explained as a sudden change of dust extinction in the line-of-sight \citep{bouvier2013}. Similarly, \citet{lamzin2017} argue that the dimmings of the classical T Tauri star RW\,Aur\,A are caused by a dust cloud, and that the behavior of the star resembles UXors, however the duration and the amplitude of the eclipses are much larger. In another case, the dimming of V409~Tau in 2011 was interpreted as occultation by a high-density feature in the circumstellar disk located $>$8\,au from the star \citep{rodriguez2015}. Recent images of the disk of AA\,Tau by the ALMA interferometer \citep{loomis2017} revealed a multi-ring structure where the rings are possibly interconnected by streamers. Such orbiting non-axisymmetric features could have a role in causing the eclipses. They may arise from hydrodynamic effects, but may also point to the presence of a planet.

If V582\,Aur is similar to the above variable objects, it possibly exhibited  deep eclipses already during its quiescent phase. Moreover, taking into account the more edge-on than face-on geometry of the disk, as well as that dust grains recondense in the vicinity of the star after an outburst, we speculate that also shorter timescale variability, like e.g. the AA~Tau phenomenon \citep[dippers,][]{bouvier2007}, may be present in quiescence. In the lack of sufficient photometric coverage of the pre-outbursts period these predictions can only be verified when the system returns to its quiescent phase. At the moment there is no indication of such a return, as the accretion rate during the whole studied period was remarkably constant (Fig.~\ref{fig:accdisk}), and also the strength of the stellar wind was unchanged (Sect.~\ref{sec:optspec}).

\section{Summary}
\label{sec:con}

We performed a detailed analysis of the optical--infrared variability of the FU Ori-type young eruptive star V582\,Aur. Our main results are the following:
\begin{itemize}
\item[--] based on archival photograpic plates at Konkoly Observatory we confirmed that the initial brightening of the source occurred between 1985 October and 1987 January, and the outburst is still ongoing with a constant, fairly high accretion rate;
\item[--] the progenitor was likely a typical classical T Tauri star, whose pre-outburst SED suggests only a moderate interstellar reddening;
\item[--] the mass of an unresolved circumstellar structure, probably a disk, is 0.04\,M$_{\odot}$;
\item[--] the long-term light curve of the object exhibited two deep minima, one in 2012 and one is still ongoing since 2016;
\item[--] the optical and near-infrared medium resolution spectra exhibit FUor-like characteristics: hydrogen and metallic lines, as well as the CO bandhead in absorption,  
\item[--] based on near-infrared photometric colors, on the comparison of multiepoch near-infrared spectra, and on fitting the multiepoch SEDs with a reddened accretion disk model, we found that both dimming events were exclusively caused by increased extinction (up to $A_V$=12.5\,mag) toward the star, while the scattered light component was unaffected;
\item[--] the obscuring dust clump, if orbiting the star, has an orbital radius of 2.8\,au, a size of $\sim$3.5\,au, and a mass of 0.004 M$_{\oplus}$, that is only a small amount of material compared to the mass accreting onto the star during the FUor outburst;
\item[--] the origin of the dust clump may be structured disk morphology with non-axisymmetric features. They may arise from hydrodynamic processes, but may also point to the presence of a planet. Alternatively, an infalling giant super-comet may also be responsible for the observed variability;
\item[--] the relationship between the obscuring disk structure(s) and the outburst remains unknown, although the relatively large distance of the dust clump from the star suggests that it is situated outside the actively accreting zone, and might have been present already in the quiescent phase. Whether it could have a role in triggering the outburst is unclear yet. The existence of such non-axysymmetric structures in the inner disk of a young low-mass star may affect the planetesimal formation process in the terrestrial zone.
\end{itemize}

\begin{acknowledgements}
The authors thank the referee for the useful comments which helped to 
improve the manuscript.
This work was supported by the Momentum grant of the MTA CSFK
Lend\"ulet Disk Research Group, the Lend\"ulet grant LP2012-31 of the
Hungarian Academy of Sciences. The William Herschel Telescope and its
service programme are operated on the island of La Palma by the Isaac
Newton Group in the Spanish Observatorio del Roque de los Muchachos of
the Instituto de Astrof\'\i{}sica de Canarias. This work is based in
part on observations made with the Telescopio Carlos Sanchez operated
on the island of Tenerife by the Instituto de Astrof\'\i{}sica de
Canarias in the Observatorio del Teide. The authors wish to thank the
telescope manager A.~Oscoz, the support astronomers and telescope
operators for their help during the observations, as well as the
service mode observers. Based on observations made with the Gran
Telescopio Canarias (GTC), installed in the Spanish Observatorio del
Roque de los Muchachos of the Instituto de Astrof\'\i{}sica de
Canarias, in the island of La Palma. We also thank Gy. Szab\'o, 
B. Cs\'ak, and Zs. Szab\'o for their assistance in scanning the 
photometric plates. This project has been supported by the GINOP-2.3.2-15-2016-00003 grant of the Hungarian National Research, Development and Innovation Office (NKFIH).  
The authors acknowledge the Hungarian National Research, Development and Innovation Office grants K-109276, K-113117, K-115709, and PD-116175. K.~V., L.~M., and \'A.~S. were supported by the Bolyai J\'anos Research Scholarship of the Hungarian Academy of Sciences.
G.H. acknowledges support from the Graduate Student Exchange
Fellowship Program between the Institute of Astrophysics of the
Pontif\'icia Universidad Cat\'olica de Chile and the
{\it Zentrum f\"ur Astronomie} of the University of Heidelberg,
funded by the Heidelberg Center in Santiago de Chile and the
{\it Deutscher Akademischer Austauschdienst (DAAD)},
by the Ministry for the Economy, Development, and Tourism’s
Programa Iniciativa Milenio through grant IC120009;
by Proyecto Basal PFB-06/2007; and by CONICYT-PCHA/Doctorado
Nacional grant 2014-63140099.
\end{acknowledgements}

\software{IRAF (http://iraf.noao.edu), GILDAS (https://www.iram.fr/IRAMFR/GILDAS/)}


\appendix

\startlongtable
\begin{deluxetable*}{lcllllllll}
\tablecaption{Optical and near-infrared photometry in magnitudes for
	  V582~Aur. Numbers in parentheses give the formal uncertainty of the last digit. \label{tab:phot}} 
\tablewidth{20cm}
\tablehead{
\colhead{Date} & \colhead{JD$\,{-}\,$2,400,000} & \colhead{$B$} &
\colhead{$V$} & \colhead{$R$} & \colhead{$I$} & \colhead{$J$} &
\colhead{$H$} & \colhead{$K_S$} & \colhead{Telescope}}
\startdata
\multicolumn{10}{c}{Photographic plate measurements} \\
\hline
1983-11-06 & 45645.51    &            & $>$18.5\tablenotemark{a}            &  &	    &   &   &   & Schmidt    \\
1985-10-21 & 46359.59	 &	      & $>$16.3\tablenotemark{a}            &  &	    &   &   &   & Schmidt    \\
1987-01-24 & 46820.38	 &	      & 13.85(15)\tablenotemark{a}          &  &	    &   &   &   & Schmidt    \\
1990-08-28 & 48132.58    &            & 14.00(20)\tablenotemark{b}          &  &	    &   &   &   & Schmidt    \\
\hline
\multicolumn{10}{c}{CCD measurements} \\
\hline
2010-09-20 & 55459.61 & 15.97(2)  & 14.24(3)  & 13.15(2) & 11.87(2) &          &          &          & Schmidt \\
2010-09-20 & 55459.70 & 	  &	      & 	 &	    & 9.72(2)  & 8.63(2)  & 7.92(2)  & TCS     \\
2010-09-21 & 55460.59 & 15.99(1)  & 14.25(1)  & 13.14(1) & 11.87(1) &          &          &          & Schmidt \\
2010-09-21 & 55460.70 & 	  &	      & 	 &	    & 9.74(2)  & 8.63(2)  & 7.84(2)  & TCS     \\
2010-09-24 & 55463.56 &           & 14.24(2)  & 13.08(4) & 11.82(3) &          &          &          & Schmidt \\
2010-09-25 & 55464.57 &           &           & 13.12(3) & 11.76(9) &          &          &          & Schmidt \\
2010-10-20 & 55478.52 & 15.82(1)  & 14.13(1)  & 13.04(1) & 11.80(1) &	       &	  &	     & Schmidt \\
2010-11-17 & 55517.51 & 16.05(1)  & 14.31(1)  & 13.21(1) & 11.93(1) &	       &	  &	     & IAC-80 \\
2010-11-17 & 55518.49 & 16.07(2)  & 14.33(1)  & 13.24(1) & 11.96(1) &	       &	  &	     & IAC-80 \\
2010-11-18 & 55518.70 & 	  &	      & 	 &	    & 9.78(2)  & 8.66(2)  & 7.92(2)  & TCS     \\
2010-11-19 & 55519.72 & 	  &	      & 	 &	    & 9.72(2)  & 8.67(2)  & 7.89(2)  & TCS     \\
2011-03-11 & 55632.24 & 15.82(3)  & 14.18(2)  & 13.11(2) & 11.85(1) &	       &	  &	     & Schmidt \\
2011-03-12 & 55633.33 &           & 14.14(1)  & 13.08(1) & 11.83(1) &	       &	  &	     & Schmidt \\
2011-03-13 & 55634.33 &           & 14.13(2)  & 13.06(1) & 11.83(1) &	       &	  &	     & Schmidt \\
2011-03-24 & 55645.34 & 15.74(2)  & 14.05(1)  & 12.99(1) & 11.75(1) &	       &	  &	     & Schmidt \\
2011-04-26 & 55678.38 &           & 14.02(1)  & 12.95(1) &	    &          &	  &	     & Schmidt \\
2011-04-26 & 55678.41 & 	  &	      & 	 &	    & 9.67(2)  & 8.59(2)  & 7.90(2)  & TCS     \\
2011-09-21 & 55826.48 & 15.72(2)  & 14.09(1)  & 13.06(1) & 11.82(1) &	       &	  &	     & Schmidt \\
2011-09-26 & 55830.57 & 15.84(1)  & 14.17(1)  & 13.10(1) & 11.87(1) &	       &	  &	     & RCC     \\
2011-09-27 & 55831.56 & 15.86(1)  & 14.16(1)  & 13.10(1) & 11.87(1) &	       &	  &	     & RCC     \\
2011-10-08 & 55842.72 & 	  &	      & 	 &	    & 9.78(2)  & 8.67(2)  & 7.92(2)  & TCS     \\
2011-10-09 & 55843.63 & 	  &	      & 	 &	    & 9.78(2)  & 8.66(2)  & 9.73(2)  & TCS     \\
2011-11-16 & 55882.49 & 16.36(1)  & 14.56(1)  & 13.45(1) & 12.15(1) &	       &	  &	     & Schmidt \\
2011-11-30 & 55895.50 & 	  &	      & 	 &	    & 9.94(2)  & 8.82(2)  & 8.07(2)  & TCS     \\
2011-11-30 & 55895.56 & 16.24(2)  & 14.50(1)  & 13.41(1) & 12.14(1) &	       &	  &	     & Schmidt \\
2012-02-10 & 55968.48 & 17.84(5)  & 15.80(2)  & 14.58(3) & 13.10(2) &	       &	  &	     & Schmidt \\
2012-02-12 & 55970.33 & 17.81(3)  & 15.92(4)  & 14.63(2) & 13.19(2) &	       &	  &	     & Schmidt \\
2012-02-14 & 55971.51 & 17.82(7)  & 15.92(4)  & 14.69(2) & 13.16(2) &	       &	  &	     & Schmidt \\
2012-03-02 & 55989.41 & 	  &	      & 	 &	    & 11.00(2) & 9.52(2)  & 8.50(2)  & TCS     \\
2012-03-04 & 55991.26 & 18.05     & 16.40     & 15.27    & 13.84    &	       &	  &	     & Schmidt \\
2012-03-05 & 55992.27 & 18.33(4)  & 16.46(2)  & 15.32(1) & 13.90(2) &	       &	  &	     & Schmidt \\
2012-03-07 & 55994.29 & 18.15(4)  & 16.52(5)  & 15.39(1) & 13.99(1) &	       &	  &	     & Schmidt \\
2012-04-11 & 56029.34 &           &           & 15.59(8) & 14.24(5) &	       &	  &	     & Schmidt \\
2012-04-27 & 56045.32 & 18.59(19) & 16.76(4)  & 15.65(5) & 14.21(4) &	       &	  &	     & Schmidt \\
2012-08-20 & 56159.57 & 16.92(2)  & 15.04(1)  & 13.81(1) & 12.43(1) &	       &	  &	     & Schmidt \\
2012-08-22 & 56161.58 & 16.79(1)  & 15.00(1)  & 13.79(1) & 12.42(1) &	       &	  &	     & Schmidt \\
2012-10-10 & 56210.63 &           & 14.57(1)  & 13.44(2) & 12.12(1) &	       &	  &	     & Schmidt \\
2012-10-10 & 56210.67 & 	  &	      & 	 &	    & 9.94(2)  & 8.82(2)  & 8.10(2)  & TCS     \\
2012-10-11 & 56211.61 &           & 14.56(1)  & 13.46(1) & 12.10(1) &	       &	  &	     & Schmidt \\
2012-10-12 & 56212.76 &           & 14.51(1)  & 13.41(1) & 12.06(1) &	       &	  &	     & Schmidt \\
2012-10-13 & 56213.67 &           & 14.49(1)  & 13.39(1) & 12.17(1) &	       &	  &	     & Schmidt \\
2012-10-14 & 56214.73 &           & 14.45(1)  & 13.35(1) & 12.02(1) &	       &	  &	     & Schmidt \\
2014-02-23 & 56712.25 & 16.19(5)  & 14.47(1)  & 13.37(1) & 12.13(1) &	       &	  &	     & Schmidt \\
2014-02-24 & 56713.25 &           & 14.45     & 13.39    & 12.14    &	       &	  &	     & Schmidt \\
2014-03-01 & 56718.32 & 16.16(6)  & 14.45(1)  & 13.37(1) & 12.13(2) &	       &	  &	     & Schmidt \\
2014-03-14 & 56731.37 & 16.48(5)  & 14.67(2)  & 13.56(2) & 12.24(1) &	       &	  &	     & Schmidt \\
2014-04-07 & 56755.32 & 16.42(10) & 14.60(2)  & 13.46(1) & 12.16(1) &	       &	  &	     & Schmidt \\
2015-01-06 & 57028.52 & 16.23(8)  & 14.39(5)  & 13.36(2) & 12.07(2) &	       &	  &	     & Schmidt \\
2015-01-06 & 57029.50 & 16.22(3)  & 14.43(2)  & 13.35(1) & 12.08(1) &	       &	  &	     & Schmidt \\
2015-01-13 & 57036.45 & 16.32(1)  & 14.53(1)  & 13.43(1) & 12.16(1) &	       &	  &	     & Schmidt \\
2015-01-16 & 57039.44 & 16.20(4)  & 14.44(1)  & 13.34(1) & 12.09(1) &	       &	  &	     & Schmidt \\
2015-01-28 & 57051.46 & 16.25     & 14.54     & 13.37    & 12.13    &	       &	  &	     & Schmidt \\
2015-02-08 & 57062.37 & 16.19(2)  & 14.38(6)  & 13.38(1) & 12.19(4) &	       &	  &	     & Schmidt \\
2015-02-16 & 57070.35 & 16.32(1)  & 14.52(2)  & 13.42(1) & 12.14(1) &	       &	  &	     & Schmidt \\
2015-03-06 & 57088.27 & 16.45(5)  & 14.68(5)  & 13.56(4) & 12.28(4) &	       &	  &	     & Schmidt \\
2015-03-08 & 57090.28 & 16.56(2)  & 14.81(1)  & 13.68(1) & 12.37(1) &	       &	  &	     & Schmidt \\
2015-03-09 & 57091.28 & 16.62(2)  & 14.84(1)  & 13.69(1) & 12.40(1) &	       &	  &	     & Schmidt \\
2015-03-16 & 57098.27 & 16.68(3)  & 14.88(3)  & 13.76(2) & 12.45(1) &	       &	  &	     & Schmidt \\
2015-04-07 & 57120.27 & 16.56(7)  & 14.83(2)  & 13.71(1) & 12.38(1) &	       &	  &	     & Schmidt \\
2015-04-08 & 57121.31 & 16.66(2)  & 14.83(1)  & 13.71(1) & 12.41(1) &	       &	  &	     & Schmidt \\
2015-04-12 & 57125.28 & 17.04(23) & 14.98(8)  & 13.84(2) & 12.52(3) &	       &	  &	     & Schmidt \\
2015-09-14 & 57279.57 & 16.89(9)  & 15.14(2)  & 13.93(2) & 12.59(3) &	       &	  &	     & Schmidt \\
2015-10-31 & 57326.54 &           & 14.96(3)  & 13.84(1) & 12.50(1) &	       &	  &	     & Schmidt \\
2015-11-10 & 57337.45 &           & 15.15(2)  & 13.98(2) & 12.64(2) &	       &	  &	     & Schmidt \\
2015-11-24 & 57351.36 & 16.98(3)  & 15.21(1)  & 14.05(2) & 12.67(1) &	       &	  &	     & Schmidt \\
2015-12-13 & 57370.26 & 16.61(1)  & 14.86(3)  & 13.67(3) & 12.40(1) &	       &	  &	     & Schmidt \\
2015-12-27 & 57383.55 & 16.68(1)  & 14.90(2)  & 13.80(2) & 12.47(2) &	       &	  &	     & Schmidt \\
2016-03-09 & 57457.28 & 18.00(5)  & 15.91(8)  & 14.71(2) & 13.25(2) &	       &	  &	     & Schmidt \\
2016-03-31 & 57479.29 & 18.48(3)  & 16.88(3)  & 15.75(5) & 14.28(4) &	       &	  &	     & Schmidt \\
2016-04-12 & 57491.28 &           & 16.78(10) & 15.68(5) & 14.35(4) &	       &	  &	     & Schmidt \\
2016-11-21 & 57714.43 & 18.43(7)  & 16.49(1)  & 15.25(2) & 13.69(1) &	       &	  &	     & Schmidt \\
2016-11-24 & 57717.49 & 18.34(1)  & 16.66(1)  & 15.33(6) & 13.80(2) &	       &	  &	     & Schmidt \\
2016-11-25 & 57717.54 &           &           &          &          & 11.08(2) & 9.62(2)  & 8.58(2)  & TCS     \\
2016-11-26 & 57719.30 & 18.65(6)  & 16.75(2)  & 15.49(2) & 13.92(1) &	       &	  &	     & Schmidt \\
2016-11-27 & 57719.56 &           &           &          &          & 11.13(2) & 9.63(2)  & 8.59(2)  & TCS     \\
2016-11-27 & 57720.30 & 18.65(4)  & 16.80(2)  & 15.53(2) & 13.96(1) &	       &	  &	     & Schmidt \\
2016-11-28 & 57721.30 & 18.59(8)  & 16.77(1)  & 15.51(1) & 13.98(2) &	       &	  &	     & Schmidt \\
2016-11-29 & 57722.45 & 18.65(2)  & 16.82(2)  & 15.59(1) & 14.00(1) &	       &	  &	     & Schmidt \\
2016-12-04 & 57727.46 & 18.66(5)  & 16.86(2)  & 15.61(1) & 14.03(1) &	       &	  &	     & Schmidt \\
2016-12-04 & 57727.67 &           &           &          &          & 11.18(2) & 9.68(2)  & 8.60(2)  & TCS     \\
2016-12-05 & 57728.68 &           &           &          &          & 11.19(2) & 9.68(2)  & 8.61(2)  & TCS     \\
2016-12-12 & 57735.34 & 18.85     & 16.67     & 15.59(1) & 14.05(2) &	       &	  &	     & Schmidt \\
2016-12-14 & 57737.30 & 18.39(27) & 16.57(4)  & 15.57(4) & 14.07(2) &	       &	  &	     & Schmidt \\
2016-12-15 & 57737.55 &           &           &          &          & 11.28(2) & 9.74(2)  & 8.65(2)  & TCS     \\
2016-12-15 & 57738.47 & 18.98(19) & 16.74(9)  &	         & 14.17(7) &	       &	  &	     & Schmidt \\
2016-12-16 & 57738.58 &           &           &          &          & 11.29(2) & 9.73(2)  & 8.62(2)  & TCS     \\
2016-12-16 & 57739.35 & 18.79(8)  & 16.83(3)  & 15.65(1) & 14.13(1) &	       &	  &	     & Schmidt \\
2016-12-17 & 57740.33 & 18.83(18) & 16.83(2)  & 15.60(5) & 14.07(3) &	       &	  &	     & Schmidt \\
2016-12-18 & 57741.46 & 18.67(10) & 16.86(3)  & 15.62(2) & 14.12(1) &	       &	  &	     & Schmidt \\
2016-12-20 & 57743.44 & 18.64(2)  & 16.92(1)  & 15.70(1) & 14.18(1) &	       &	  &	     & Schmidt \\
2016-12-21 & 57744.34 & 18.69(2)  & 16.91(2)  & 15.72(1) & 14.20(1) &	       &	  &	     & Schmidt \\
2016-12-28 & 57751.37 & 18.59(8)  & 16.90(3)  & 15.78(1) & 14.27(2) &	       &	  &	     & Schmidt \\
2016-12-29 & 57752.39 & 18.71(4)  & 16.97(2)  & 15.80(1) & 14.33(1) &	       &	  &	     & Schmidt \\
2016-12-30 & 57753.39 & 18.73(5)  & 16.94(1)  & 15.78(2) & 14.32(2) &	       &	  &	     & Schmidt \\
2016-12-31 & 57754.41 & 18.66(3)  & 16.92(2)  & 15.77(1) & 14.32(1) &	       &	  &	     & Schmidt \\
2017-01-01 & 57755.40 & 18.60(7)  & 16.90(1)  & 15.78(2) & 14.32(1) &	       &	  &	     & Schmidt \\
2017-01-02 & 57756.23 & 18.66(7)  & 16.90(2)  & 15.75(2) & 14.29(1) &	       &	  &	     & Schmidt \\
2017-01-05 & 57759.39 &           &           &          &          & 11.50(2) & 10.02(2)\tablenotemark{c} & 8.52(2)\tablenotemark{d}  & LIRIS   \\
2017-01-10 & 57763.64 &           &           &          &          & 11.61(2) & 9.92(2)  & 8.70(2)  & TCS     \\
2017-01-11 & 57764.57 &           &           &          &          & 11.61(2) & 9.93(2)  & 8.72(2)  & TCS     \\
2017-01-11 & 57765.29 &		  & 16.96(1)  & 15.76(3) & 14.36(1) &	       &	  &	     & Schmidt \\
2017-01-19 & 57773.38 & 18.79(7)  & 17.03(3)  & 15.95(1) & 14.56(2) &	       &	  &	     & Schmidt \\
2017-01-20 & 57774.43 & 18.65(7)  & 16.99(2)  & 15.96(2) & 14.59(1) &	       &	  &	     & Schmidt \\
2017-01-22 & 57776.41 & 18.65(4)  & 17.03(2)  & 15.95(2) & 14.60(1) &	       &	  &	     & Schmidt \\
2017-01-23 & 57777.21 & 18.77(4)  & 17.04(1)  & 15.98(1) & 14.61(1) &          &	  &	     & Schmidt \\
2017-01-24 & 57778.35 & 18.74(5)  & 17.06(3)  & 15.97(2) & 14.62(1) &          &	  &	     & Schmidt \\
2017-01-25 & 57778.59 &           &           &          &          & 11.78(2) & 10.04(2) & 8.79(2)  & TCS     \\
2017-01-26 & 57780.25 & 18.62(5)  & 17.00(2)  & 15.97(3) & 14.61(2) &          &	  &	     & Schmidt \\
2017-01-27 & 57781.24 & 18.75(9)  & 17.05(2)  & 15.98(1) & 14.63(1) &          &	  &	     & Schmidt \\
2017-01-28 & 57782.28 & 18.65(2)  & 17.02(3)  & 15.99(1) & 14.65(2) &          &	  &	     & Schmidt \\
2017-01-29 & 57783.34 & 18.70(4)  & 17.05(2)  & 16.02(3) & 14.68(2) &          &	  &	     & Schmidt \\
2017-02-03 & 57788.45 &           &           &          &          & 11.85(2) & 10.08(2) & 8.83(2)  & TCS     \\
2017-02-05 & 57789.56 &           &           &          &          & 11.86(2) & 10.08(2) & 8.81(2)  & TCS     \\
2017-02-06 & 57791.36 &           &           &          &          & 11.84(2) & 10.28(2)\tablenotemark{c} & 8.64(2)\tablenotemark{d}  & LIRIS   \\
2017-02-10 & 57795.32 & 18.49(9)  & 16.99(4)  & 16.02(2) & 14.69(1) &          &	  &	     & Schmidt \\
2017-02-11 & 57796.25 & 18.66(20) & 17.00(3)  &          & 14.72(1) &          &	  &	     & Schmidt \\
2017-02-12 & 57797.23 & 18.70(5)  & 17.08(2)  & 16.04(1) & 14.73(1) &          &	  &	     & Schmidt \\
2017-02-13 & 57798.43 & 18.61(17) & 17.08(5)  & 16.03(2) & 14.71(1) &          &	  &	     & Schmidt \\
2017-02-15 & 57800.40 & 18.72(8)  & 17.09(2)  & 16.05(2) & 14.74(1) &          &	  &	     & Schmidt \\
2017-02-23 & 57808.37 & 18.77(6)  & 17.08(4)  & 16.07(2) & 14.70(11)&          &	  &	     & Schmidt \\
2017-02-24 & 57808.50 &           &           &          &          & 11.85(2) & 10.07(2) & 8.85(2)  & TCS     \\
2017-03-01 & 57814.27 & 18.78(9)  & 17.09(5)  & 15.96(1) & 14.50(7) &          &	  &	     & Schmidt \\
2017-03-02 & 57815.32 & 18.65(2)  & 17.04(2)  & 15.97(1) & 14.50(1) &          &	  &	     & Schmidt \\
2017-03-03 & 57816.24 & 18.57(12) & 16.99(2)  & 15.88(2) & 14.47(2) &          &	  &	     & Schmidt \\
2017-03-04 & 57817.24 & 18.58(22) & 17.12(10) & 15.87(3) & 14.46(2) &          &	  &	     & Schmidt \\
2017-03-08 & 57821.40 & 19.20     & 16.97(13) & 15.89(2) & 14.43    &          &	  &	     & Schmidt \\
2017-03-10 & 57823.28 & 18.48(1)  & 16.98(4)  & 15.89(3) & 14.48(2) &          &	  &	     & Schmidt \\
2017-03-14 & 57827.30 & 18.75(9)  & 17.00(7)  & 15.93(1) & 14.51(2) &          &	  &	     & Schmidt \\
2017-03-15 & 57828.33 & 18.67(7)  & 17.06(2)  & 15.96(2) & 14.52(1) &          &	  &	     & Schmidt \\
2017-03-25 & 57838.27 & 18.80(8)  & 17.15(5)  & 16.00(2) & 14.55(1) &          &	  &	     & Schmidt \\
2017-03-28 & 57841.32 & 18.79(4)  & 17.09(2)  & 15.96(1) & 14.49(1) &          &	  &	     & Schmidt \\
           & 57841.50 &           &           &          &          & 11.48(2) &          & 8.66(2)  & TCS     \\
2017-04-01 & 57845.28 & 18.62(7)  & 16.95(2)  & 15.88(1) & 14.41(3) &          &	  &	     & Schmidt \\
2017-04-03 & 57847.31 &           & 16.75(4)  & 15.81(2) & 14.32(2) &          &	  &	     & Schmidt \\
           & 57851.44 &           &           &          &          & 11.34(2) & 9.73(2)  & 8.62(2)  & TCS     \\
2017-04-09 & 57853.31 & 18.37(14) & 16.87(8)  & 15.70(1) & 14.17(2) &          &	  &	     & Schmidt \\
           & 57856.44 &           &           &          &          & 11.25(2) & 9.67(2)  & 8.53(4)  & TCS     \\
2017-04-13 & 57857.29 & 18.69     & 17.02(4)  & 15.79(2) & 14.20(3) &          &	  &	     & Schmidt \\
           & 57857.44 &           &           &          &          & 11.26(2) & 9.66(2)  & 8.52(2)  & TCS     \\
2017-04-14 & 57858.28 & 18.27     & 16.96(4)  & 15.70(3) & 14.19(4) &          &	  &	     & Schmidt \\
2017-04-16 & 57860.29 & 19.03     & 17.03(5)  & 15.78(2) & 14.18(1) &          &	  &	     & Schmidt \\
           & 57860.41 &           &           &          &          & 11.27(2) & 9.68(2)  & 8.58(2)  & TCS     \\
2017-04-17 & 57861.29 & 18.11     & 16.89(2)  & 15.72(2) & 14.17(3) &          &	  &	     & Schmidt \\
2017-04-24 & 57868.30 & 19.06     & 16.97(1)  & 15.83(2) & 14.32(2) &          &	  &	     & Schmidt \\
2017-08-24 & 57989.58 & 18.22(3)  & 16.41(3)  & 15.17(2) & 13.65(1) &          &	  &	     & Schmidt \\
2017-09-01 & 57997.58 & 18.45(6)  & 16.54(2)  & 15.30(1) & 13.75(1) &          &	  &	     & Schmidt \\
2017-09-23 & 58019.55 & 18.59(5)  & 16.82(2)  & 15.60(2) & 14.02(1) &          &	  &	     & Schmidt \\
2017-09-27 & 58023.54 &           & 16.63     & 15.39(6) & 13.84(3) &          &	  &	     & Schmidt \\
2017-09-28 & 58024.55 & 18.55(7)  & 16.75(2)  & 15.45(1) & 13.88(1) &          &	  &	     & Schmidt \\
2017-10-02 & 58028.58 & 18.58(5)  & 16.74(6)  & 15.46(3) & 13.89(1) &          &	  &	     & Schmidt \\
2017-10-11 & 58038.45 & 18.53(12) & 16.70(2)  & 15.39(2) & 13.81(2) &          &	  &	     & Schmidt \\
2017-10-14 & 58040.58 & 18.54(4)  & 16.58(2)  & 15.33(3) & 13.74(2) &          &	  &	     & Schmidt \\
2017-10-15 & 58041.52 & 18.44(3)  & 16.62(4)  & 15.35(1) & 13.77(1) &          &	  &	     & Schmidt \\
2017-11-06 & 58064.36 & 17.90(14) & 15.91(5)  & 14.66(5) & 13.19(2) &          &    &      & Schmidt \\
2017-11-24 & 58082.39 & 18.12(2)  & 16.17(2)  & 14.88(1) & 13.36(1) &          &    &      & Schmidt \\
2017-12-02 & 58089.58 & 17.79(3)  & 16.02(2)  & 14.76(2) & 13.24(1) &          &    &      & Schmidt \\
2017-12-05 & 58092.68 &           & 15.85(4)  & 14.72(4) & 13.23(3) &          &    &      & Schmidt \\
2017-12-07 & 58095.36 & 17.79(3)  & 15.99(2)  & 14.73(2) & 13.21(1) &          &    &      & Schmidt \\
\hline
\enddata
\tablenotetext{a}{103-D emulsion, GG14 filter}
\tablenotetext{b}{103-D emulsion, no filter}
\tablenotetext{c}{LIRIS H$_c$ filter}
\tablenotetext{d}{LIRIS K$_c$ filter}
\end{deluxetable*}


\end{document}